\documentclass[journal]{IEEEtran}
\ifCLASSINFOpdf
\else
\fi
\usepackage{dblfloatfix}
\usepackage{url}

\usepackage{svg}
\usepackage{amsmath}
\usepackage{pdfpages}
\usepackage{hyperref}
\usepackage{url}
\usepackage{cite}
\usepackage{algorithm}
\usepackage{algpseudocode}
\usepackage{balance}

\algnewcommand\algorithmicforeach{\textbf{for each}}
\algdef{S}[FOR]{ForEach}[1]{\algorithmicforeach\ #1\ \algorithmicdo}
\makeatletter
\newenvironment{megaalgorithm}[1][htb]{
    \renewcommand{\ALG@name}{\textbf{Algorithm}}
   \begin{algorithm}[#1]
  }{\end{algorithm}}
  
\algnewcommand\algorithmicswitch{\textbf{switch}}
\algnewcommand\algorithmiccase{\textbf{case}}
\algnewcommand\algorithmicassert{\texttt{assert}}
\algnewcommand\Assert[1]{\State \algorithmicassert(#1)}%
\algdef{SE}[SWITCH]{Switch}{EndSwitch}[1]{\algorithmicswitch\ #1\ \algorithmicdo}{\algorithmicend\ \algorithmicswitch}%
\algdef{SE}[CASE]{Case}{EndCase}[1]{\algorithmiccase\ #1}{\algorithmicend\ \algorithmiccase}%
\algtext*{EndSwitch}%
\algtext*{EndCase}%
\usepackage{graphicx}
\usepackage{makecell}
\graphicspath{ {./images/} }
\begin{document}
%
\title{B-DAC: A Decentralized Access Control Framework on Northbound Interface for Securing SDN Using Blockchain}
%
%
\author{\IEEEauthorblockN{Phan The Duy\IEEEauthorrefmark{1}\IEEEauthorrefmark{2},
Hien Do Hoang\IEEEauthorrefmark{1}\IEEEauthorrefmark{2}, Do Thi Thu Hien\IEEEauthorrefmark{1}\IEEEauthorrefmark{2}, Anh Gia-Tuan Nguyen\IEEEauthorrefmark{1}\IEEEauthorrefmark{2} and Van-Hau Pham\IEEEauthorrefmark{1}\IEEEauthorrefmark{2}}\\
\IEEEauthorblockA{\IEEEauthorrefmark{1}Information Security Laboratory, University of Information Technology, Ho Chi Minh city, Vietnam\\
\IEEEauthorrefmark{2}Vietnam National University, Ho Chi Minh city, Vietnam\\
\{duypt, hiendh, hiendtt, anhngt, haupv\}@uit.edu.vn}}

%
%

\markboth{Journal of \LaTeX\ Class Files, August~2021}%
{Shell \MakeLowercase{\textit{et al.}}: Bare Demo of IEEEtran.cls for IEEE Journals}
%



\maketitle



%
\IEEEpeerreviewmaketitle

\begin{abstract}
Software-Defined Network (SDN) is a new arising terminology of network architecture with outstanding features of orchestration by decoupling the control plane and the data plane in each network element. Even though it brings several benefits, SDN is vulnerable to a diversity of attacks.  Abusing the single point of failure in the SDN controller component, hackers can shut down all network operations. More specifics, a malicious OpenFlow application can access to SDN controller to carry out harmful actions without any limitation owing to the lack of the access control mechanism as a standard in the Northbound. The sensitive information about the whole network such as network topology, flow information, and statistics can be gathered and leaked out. Even worse, the entire network can be taken over by the compromised controller. Hence, it is vital to build a scheme of access control for SDN’s Northbound. Furthermore, it must also protect the data integrity and availability during data exchange between application and controller. To address such limitations, we introduce B-DAC, a blockchain-based framework for decentralized authentication and fine-grained access control for the Northbound interface to assist administrators in managing and protecting critical resources. With strict policy enforcement, B-DAC can perform decentralized access control for each request to keep network applications under surveillance for preventing over-privileged activities or security policy conflicts. To demonstrate the feasibility of our approach, we also implement a prototype of this framework to evaluate the security impact, effectiveness, and performance through typical use cases.
\end{abstract}
\section{Introduction}\label{sec1}
\IEEEPARstart{S}{}oftware-Defined Network (SDN) is an emerging technology that has been gaining significant attention among IT professionals and the public. One of the reasons for this interest is the rising prominence of network management with a flexible mechanism provided by the centralized controller. In fact, SDN has been considered and deployed in various real-world environments and potential cases in the near future. It includes large-scale data centers, SDN-based cloud, namely Huawei NovoDC \cite{c1}, CloudFabric \cite{c2}. Also, it  is impossible not to mention Microsoft Azure \cite{c3}, IBM Network Services \cite{c4}, Cisco \cite{c5}, NTT DOCOMO \cite{c6} in this adoption trend. In addition, witnessing the surge in the number of devices and traffic volumes, the system architectures of next-generation networks (5G/6G) are introduced based on SDN technologies \cite{c7}. Similarly, edge computing is considered as one of promising fields of SDN integration to facilitate the management and operation of edge servers and various IoT devices \cite{c8}. This can be explained that heterogeneous networks can leverage the SDN paradigm to manage numerous IoT and network devices over the large-scale network \cite{c9}.

Although SDN has great potential for the reconstruction of the future Internet, it is facing a diversity of technical challenges and security issues. To be more precise, the SDN controller becomes the most vulnerable component in SDN architecture, since it adequately manages the entire network. In this circumstance, an access control solution for the controller and its Northbound interface plays a vital role to prohibit disruption of the whole network functions from rogue network applications. Generally, the access control model is responsible for restricting the activity of network applications and enforcing security policies to protect the network from unauthorized access. Most of the existing works propose management schema laid on a specific controller. This makes it difficult and inadequate to port their models to other controllers without modifying a bunch of source codes in the SDN controller. That is why our system aims to independently operate with the controller while providing the assurance of tamper resistance in confidential artifacts. In this context, it enforces the scheme of controlling the application identity, as well as their behaviors and logging as audit trails.

Besides, when an SDN application is compromised, it can create powerful flow rules to process packets according to their intent, such as modifying the routing path, the header of a data packet. As a result, there are conflicts of flow rules in the network devices due to commands from the controller which are generated by malicious applications. Unfortunately, several attempts of implementing access control systems ignore the legality of flow rules that result in the invalidation of network functions and security issues. Thus, in addition to the AAA scheme (Authentication, Authorization, Accounting), it is necessary to detect the malicious or conflicting flow rules to verify the correctness of network operation before being inserted into OpenFlow switches.

Furthermore, previous systems for securing SDN controllers from malicious applications ignore the vulnerabilities and risks of their mechanism itself. Indeed, attackers can intrude and compromise the monitoring system to gain higher privileges by modifying the security policy database which belongs to the access control system. Consequently, the control scheme at the Northbound interface can be bypassed, and the controller can be deceived to perform harmful activities taken down the whole network system.

With the booming era of blockchain in various applications, the blockchain-based approach has gathered considerable attention from both industry and academia, especially in security problems. Its centerpiece is a decentralized structure that allows the features of assuring tamper-proof, immutable, time-stamped record keeping. Recently, many research fields witnessed an unprecedented rise in the development, integration, and maturity of blockchain adoption. It is proved that the principle of data security and privacy, as well as the trust-based relationship with personal information on the Internet, can be revolutionized by leveraging blockchain. Since the explosion of cryptocurrency, the advocates introduce a range of prototypes utilizing blockchain for enhancing security, privacy-preserving in data storing and sharing. There are such applications in the context of networking, IoT, smart environment, education, healthcare, fintech, big data, artificial intelligence, and cybersecurity, etc. \cite{c10}. With outstanding characteristics, such as decentralization, immutability, and auditability, blockchain gives a potential approach for applying to AAA systems (Authentication, Authorization, Accounting). Moreover, the previous works propose AAA systems based on the centralized model. This makes the AAA system become a Single point of Failure. Upon suffered from attacks, a security monitoring system can be interrupted or eavesdropped. As a result, sensitive data can be leaked out. Hence, it is important to find a novel solution to overcome these weaknesses. To resolve such challenges, blockchain-based approaches have rapidly been considered as a potential solution for access control systems to ensure the integrity and tamper resistance of sensitive data.

Our proposed solution is first motivated by Trust Trident \cite{c11}, a controller-independent approach for authentication framework. Then, we continue to integrate blockchain features into the access control system on the Northbound interface in a fine-grained manner, named B-DAC. It is designed to be a decentralized framework for handling network applications in securing controller operation. It tackles the issues of security, privacy, scalability, and reliability in the scheme of traditional authentication. Our framework is independent of the controller so that this mechanism can easily adapt to the other specific controllers. Specifically, our access control system is responsible for authenticating, authorizing, and monitoring applications in communication with the controller. In addition, flow rules are validated to prevent conflicts before being installed into the switches according to the command of the controller through network applications.

The rest of this paper covers the following sections. \textbf{Section~\ref{sec2}} gives an overview of SDN, security issues on Northbound interface. Problem statement and research goal are discussed on \textbf{Section~\ref{sec3}}. Consequently, the architecture and main operation principle of our B-DAC framework, are described in \textbf{Section~\ref{sec4}}. An example of request processing flow in our B-DAC is presented in \textbf{Section~\ref{sec5}}. Later, in \textbf{Section~\ref{sec6}}, we present a prototype implementation and evaluation of our mechanism. \textbf{Section~\ref{sec7}} introduces the existing works relating to the access control framework for SDN or network applications. Additionally, we also indicate that it is potential to apply blockchain in many security solutions, which complies with the requirement of decentralization, tamperproof and integrity, infeasible hacking. After that, in \textbf{Section~\ref{sec8}}, we draw conclusions, also further discuss our prototype system and future work.
\section{Background}
\label{sec2}
This section introduces the background on SDN, including the basic architecture and its security problems. Also, we give an overview of the main concepts and principles of blockchain technology, then discuss the possibility of utilizing blockchain in the context of SDN security.
\subsection{Overview of SDN}
\begin{figure}[!b]
\centering
\includegraphics[width=0.4\textwidth]{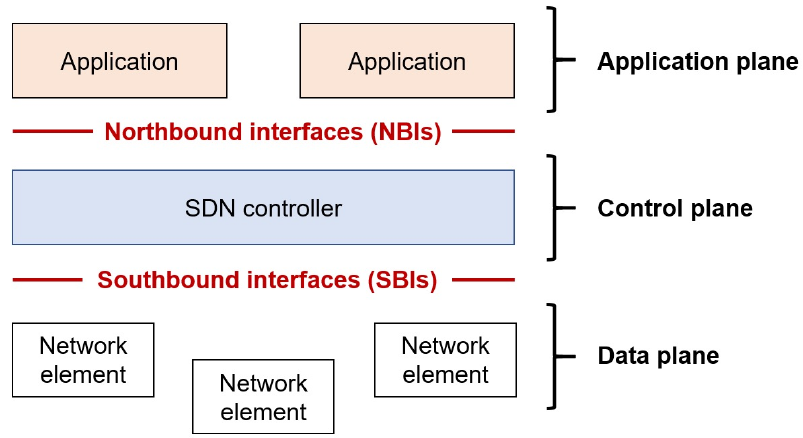}
\caption{Overview of SDN architecture}
\label{fig1}
\end{figure}

In contrast to the traditional networks, Software-defined network (SDN) adequately changes the network architecture by separating the network logic from the underlying forwarding devices. The three-layer architecture of SDN is shown in \textbf{Fig.~\ref{fig1}}. It consists of the application plane, the control plane, and the data plane. Concerning communication links, the Northbound interface is set as the connection between the controller and applications, whereas the controller and the physical networking hardware communicate to each other via the Southbound interface.

In SDN, the controller – a core component placed in the control plane, plays as the brain of the network. It takes the duty of directing traffic to desired destinations. The routing information is then delivered to the network devices on the data plane to forward the traffic accordingly via flow rule installation. The process of exchanging messages between the control and data plane is going under the support of OpenFlow protocol – a standard Southbound interface API. This centralized control manner not only provides the whole view of the network but also allows to easily program, modify and manage the network configuration. In fact, we do not have to access to each network device to reconfigure it. The third-party applications can leverage the management information in the controller to perform various operations, such as load balancing or statistics. To access and utilize network resources, these applications need to use the Northbound interface as the intermediate channel to communicate with the controller.

This architectural transformation rejuvenates the network layer, allowing centralized management and programmability of the networks according to the flexible security policy. More specifics, anyone can develop and deploy an application (firewall, proxy, load balancer, etc.) to the SDN network provided that it supports OpenFlow protocol.
\subsection{Security issues from Openflow applications and Northbound interface on controller}
Being a new and rising technique, SDN may lack security mechanisms to protect itself from malicious actions and become a target of attacks, according to \cite{c12}. Also mentioned in this research, the controller and the controller-application interface are two of the critical positions that are exploitable. 

Considered as the brain of SDN, the centralized management of the controller can be deceived to perform harmful actions. When it is compromised, an attacker can easily get in his hand a lot of network information such as topology, flow rule tables, connection, and statistical information. Even worse, the network configurations can also be unauthorized manipulated to meet a specific purpose of the attacker. In this case, the correctness of the network operations will be affected. 

Along with the direct attack on the controller, an attacker can also get into it in other ways, such as using NBI. This interface allows applications to interact with the controller and its managed network. However, unlike the well-known Southbound interface (SBI) standard OpenFlow, there is a lack of a standard as well as the consideration in security improvement of this interface. Hence, if no security policy is applied, NBI can become a vulnerability in the SDN architecture, where malicious applications can have access to it and then seize control over the controller. 

In the AIM-SDN study \cite{c13}, V.H. Dixit et al. have conducted various attack scenarios targeting SDN. NBI is also described as an easily exploitable component via many attacks targeting the confidentiality, integrity, and availability of the network. For example, the integrity of network information can be damaged by using NBI as well as SBI, so that undesired traffic can be allowed to forward. Moreover, the authors also state the problem of the unbounded number of flow rules that can be installed. Specifically, the controller can unconditionally accept all flow installation requests from applications. In such circumstances, the network performance can be degraded, even worse, it can lead to resource starvation for normal operation.

Meanwhile, another work called DELTA \cite{c14} also try to investigate various attack scenarios in SDN. About 20 known attacks along with 7 unknown ones from SDN applications have been successfully reproduced and discovered. Some attacks take place in or relate to NBI. For instance, the Service-unregistration attack allows all applications to register to receive and parse control messages from switches. This feature could lead to the effect of compromised applications on the legal services of other applications. Besides that, Application Eviction attacks can cause a legitimate application to change its status from ACTIVE to another state and stop working.
\subsection{Blockchain}
The emergence of Bitcoin is the first launch of blockchain technology, which is received tremendous attention from researchers and the general community. To offer the highlight features of ensuring integrity and tamper resistance, blockchain consists of many techniques ranging from cryptography, peer-to-peer network, and consensus protocol as game theory. To start with, it leverages the power of cryptography such as hash, asymmetric cryptography, digital signatures to ensure that data are kept securely and confidentially. Regarding network models, the distributed database is the main thought in blockchain. It is a public ledger that contains all records of digital transactions transmitted among the involved parties. This ledger is replicated and stored in the whole peer-to-peer nodes. Thus, it makes transaction records accessible even if several nodes are unavailable or disrupted. Moreover, a certain consensus algorithm is used to tackle the problem of synchronization from the distributed database. The consensus process also keeps the database of the blockchain under the control of multiple nodes. A new transaction data must be validated by all nodes before being added to the blockchain. Hence, it is so far difficult, even infeasible for hackers to manipulate records of the ledger since they must have controls over multiple nodes to overcome the consensus scheme.

In the early days of blockchain, it is designed as an open and distributed ledger. With its outstanding advantages, blockchain has been applied in many fields \cite{c10} while satisfying various security and privacy requirements. In fact, many types of blockchain have been developed to meet the complicated requirements. Generally, there are three primary categories of blockchain including public blockchain, private blockchain, and hybrid blockchain. To start with, anyone can join the public blockchain as a role of user or miner, regardless of their geographic location. Therein, consensus algorithms are fully transparent with users. All transactions that take place on the public ledger can be verified by everyone. The public blockchain network encourages people to participate and reward them for joining in the consensus scheme. Meanwhile, in comparison with the public blockchain, users needs an approval to participate in the private blockchain network. Also, transactions are kept private, where only participants with given permissions can read or write transactions. Usually, a private blockchain belongs to an organization. The consortium blockchain, also named permissioned blockchain, are considered as a subset of private blockchains. Its main distinguishing characteristic is that they are governed by a group rather than an entity like private blockchain. In a private blockchain, it can process hundreds or even thousands of transactions per second, as the number of authorized participants is lesser. Combining the advantages of both private and public blockchain, the hybrid blockchain is created. By this way, it leverages the privacy of private and consortium blockchains while maintaining the security and transparency of the public blockchain. This is considered as a suitable choice for those who need to keep a portion of data public and transparent while other portions need to be private.

Also mentioned in \cite{c10}, blockchain has been applied in many fields in Industry 4.0 to resolve security and privacy issues of existing system. This is offered by its decentralized model, cryptographic security benefits and fault tolerance. In education, blockchain-based digital certificates can replace for paper or a regular digital one, for example. Cryptocurrency like Bitcoin, Ether is used in finance, banking, e-commerce for implementing trade. Blockchain also allows developers to build medical sharing applications for patients and doctors in a privacy-preserved way. Besides, in terms of security and networking, blockchain also has its first applications. With the characteristics of data persistence and decentralization, blockchain is suitable for log storage or sensitive and reliable data sharing systems. In particular, the blockchain adoptions are found in several domains such as Artificial Intelligence (AI) \cite{c15}, IoT \cite{c16} \cite{c17}, intrusion detection \cite{c18}, or SDN \cite{c19} \cite{c20}. This potential is shown that there is much room for developing applications which leverages blockchain features to enhance security of specific systems. To summarize, blockchain has now been foreseen as a trusted paradigm to keep securely sensitive data with the assurance of tamper resistance and integrity.

\subsection{Hyperledger Fabric: A permissioned blockchain for privacy-preserving, scalable and low-cost approach}
\label{HyperledgerFabric}
Private blockchains have been more in the spotlight in the industry recently because they are much faster, low-cost, and privacy-oriented compared to the public blockchain \cite{minhojo2020_blockchainIoT}, \cite{Mohammad2020_privateblockchainIot}. Among them, Hyperledger Fabric \cite{c43} is a cross-industry collaborative project hosted by Linux Foundation, aiming to fit into enterprise architecture and allow organizations to customize networking rules for various consensus protocols. Hyperledger Fabric is an open-source implementation of permissioned enterprise-grade blockchain for running smart contracts, named chaincode. It consists of three groups of nodes which are peer, orderer, and client owned by different organizations. The task of identifying all nodes is carried by a Membership Service Provider (MSP). 

\begin{figure}[!t]
\centering
\includegraphics[width=0.5\textwidth]{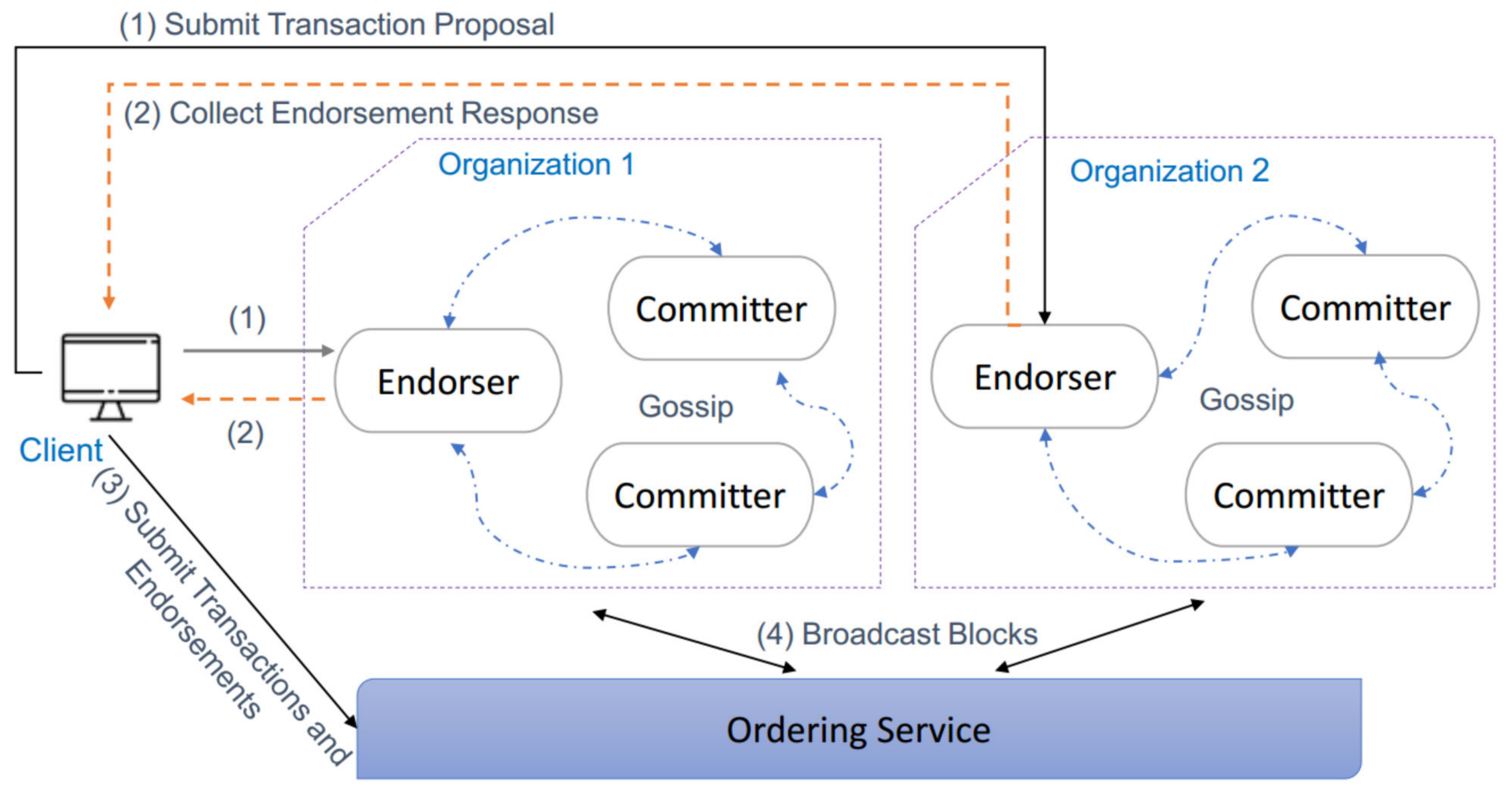}
\caption{An example of Hyperledger Fabric blockchain with two organizations}
\label{fig-hyperledger-fabric}
\end{figure}

In comparison with public blockchain such as Ethereum or Bitcoin, Hyperledger Fabric has a novel architecture, called three-phase execute-order-validate transaction flow. This architecture makes each Hyperledger Fabric's transaction must undergo three stages: endorsement execution, ordering, and validation \cite{XU2021_hyperledgerfabric}. For instance, a client first submits a transaction proposal to peer nodes assigned by the endorsement policy in advance. Next, the transaction proposal is executed on these peers by invoking an appropriate chaincode. Then, the client must collect enough endorsement results from many peers before submitting the transaction to the ordering services. The ordering service, which is located on orderer nodes, creates a total order for all transactions and builds blocks through a specific consensus algorithm. Subsequently, these blocks are broadcasted to all peers by gossip protocol for transaction verification to update the ledger state. \textbf{Fig.~\ref{fig-hyperledger-fabric}} depicts an example of Hyperledger Fabric blockchain with two organizations. Each of them has three peers: one endorser and two committers.

Regarding real-world use cases, many organizations such as enterprises in a specific field can collaborate to build the permissioned Blockchain like Hyperledger Fabric to provide a mutual decentralized platform for their operation. According to the survey of Julien Polge et al. \cite{p_blockchain_kics}, Hyperledger Fabric is also considered as the most promising platform among permissioned blockchains for industrial use cases thanks to the characteristics of privacy, latency, scalability.

In many use cases for cloud authentication and management, some approaches such as SEPSE \cite{2021_yuanzhang_blockchain_encr}, Chronos+ \cite{2021_yuanzhang_chronos} or a work of Yuan Zhang et al. \cite{2021_yuanzhang} use Ethereum, a public blockchain for implementation to provide accurate time-stamp and integrity proof for cloud services. Such approaches do not require employing miners or stakeholders for setting up the platform. It is also more secure due to the decentralization and active participation in the network. Nevertheless, public blockchain takes end users to spend a large amount of financial cost and time for transaction verification \cite{qinwang2020_blockiot}. This is where a private blockchain or consortium blockchain comes into play. But, a private blockchain is more prone to hacks, risks, and data manipulation since it is more centralized than a public one. To reduce this risk, consortium blockchain with more nodes and organizations can help. The permissioned blockchain places restrictions on who is allowed to participate in the network, and only in certain transactions. This grant control can improve the speed and throughput of transactions to cater to the enterprise's requirements. Specifically, it is getting more effective than a public blockchain in the context of Industrial Internet of Things (IIoT) applications \cite{c17}, \cite{minhojo2020_blockchainIoT} because the scalability and privacy issues can be resolved by permissioned blockchain adoption. However, the final selection of a blockchain framework for a specific case study is always a trade-off.
\section{Problem Statement}
\label{sec3}
Recently, several existing SDN authentication systems that defend SDN controllers against malicious OpenFlow applications have shown their potential effectiveness \cite{poison_app_Ujcich}. Nevertheless, they still have some fundamental problems that should be addressed to improve access control mechanisms for an SDN controller \cite{c21} \cite{c22}. In more detail, these solutions are generally controller-dependent, where source codes of the controller need to be altered for granting access permissions to external applications. Besides, the control mechanism is susceptible to denial of services or spoofing attacks itself, which breaks the entire security policy. Such incidents occur if the management system for issuing privileges is compromised. The malicious OpenFlow apps can still infect the controller for seizing network control if they easily exploit the database of security policy to perform privilege escalation. In addition, the activity logs captured by the management system can be altered to cover malicious actions. Therefore, the scheme of access control should guarantee fault tolerance, and log integrity for further investigation purposes.
\subsection{Current issues}
In this paper, we primarily focus on the following key deficiencies of existing access control systems for SDN controllers. This is to make the controller more robust and to prevent the network system from harmful actions performed by malicious network applications.
\subsubsection{Controller-dependent}
If the design of a permission-based control model is tightly coupled with a specific controller, its deployment can become a complicated task. This requires modifications in the source code of that controller to enable the access control scheme. In addition, approaches aiming to be easily implemented in various controller types should prefer the pluggable architecture instead of the all-in-one approach on controller \cite{c23} \cite{c24}. Unfortunately, most of the previous studies related to securing the Northbound interface have been strongly controller-dependent, which means they cannot be ported to other controllers. Therefore, a flexible framework is essential to monitor network applications without any dependence on a specific controller to avoid the issue of portability deficiency.
\subsubsection{Deficiencies of Authentication – Authorizing – Accounting}
To protect the SDN controller from malicious actions in the Northbound interface, a fine-grained access control system plays a crucial role to manage OpenFlow applications in consuming network resources. A rogue application allows attackers to take over the entire network by compromising the SDN controller by exploiting the Northbound API. Upon taking over, attackers can modify or inject the data flow of the network and cause serious consequences such as loss or theft of data of users and enterprises \cite{c25}. Therefore, it should provide a scheme of authentication, authorization, accounting (AAA) services to grant access permissions for OpenFlow apps to defend malicious attacks from rogue ones.

Authentication: This is one of the key features of any access control system, where participants are verified by pre-registered identity. In the context of SDN controller and OpenFlow applications, this process is important to determine which ones can interact with the controller to use the network resources. However, attackers can produce a counterfeit entity or fake program to fool the controller to achieve their illegitimate goals. For instance, after bypassing the authentication scheme, they can send malicious configurations to network devices through rule installation. Besides that, a tampered controller with IP spoofing can provide forged network resources for other genuine applications. It would lead to incorrect management and orchestration from SDN applications.

Authorization: Upon being successful in the authentication process, OpenFlow application should be given relevant grants to the network system. Specifically, every request stemming from such application is adequately taken under observation to control privileges escalation. Obviously, it is better to issue at least permissions for their task’s functions.

Accounting: We can easily get in trouble with security investigation or evaluating the trustworthiness of a specific application if there is no record of the operation history. This proves that a logging scheme is a preferred method for tracing what happens in the network relating to a certain subject.
\subsubsection{Lack of tamperproof and integrity \& Single point of failure (SPoF)}

Although there are several AAA systems for controlling applications when they communicate with SDN controllers, these proposed approaches encounter various problems \cite{c26}. Their limitations can be depicted in the issues of SPoF due to the centralized architecture of traditional access control schemes, which depend on only one server machine. Also, they lack tamperproof and integrity for the AAA database. Hackers can intrude on the system and modify the security policy, audit log, etc. In fact, they intend to cover their track in the network, which makes network forensics more difficult. In traditional access control schemes, the centralization of sensitive information in the context of IoT, cloud, or fog computing can lead to critical risks of data damage and leakage. For instance, SDN-enabled IoT with an enormous number of heterogeneous devices should not use the centralized architecture of traditional authentication mechanisms due to the issues of security, privacy, scalability, and reliability \cite{c27}. So, the vital role of availability and security-enhanced itself for access control mechanism should be taken under consideration carefully and strictly.

\subsubsection{Malicious flow rules injection}
In addition to authenticating and recording all activities of OpenFlow applications with a timestamp, monitoring systems are also required to implement flow rules verification on flow rule installation requests from applications sent to the controller. The reason for this is that different network applications can use Northbound API to deploy their intended flow rules into network elements like switches. For instance, a rogue application that has WRITE permission can insert a new rule to eavesdrop on the network traffic on the specific link. It can also remove or update existing flows on switches to bypass security checking of network firewall for further purposes, etc. Moreover, these flow rules sometimes generate conflicts or security issues in the forwarding plane. In consequence, network functions become invalid or produce incorrect or unexpected routing decisions. Hence, flow rules injection can break down operations in the network, which lacks a mechanism to control rule conflict before it takes effect at switches \cite{c28} \cite{c29}.
\subsubsection{Exhausting controller resources}
Currently, as the SDN controller is open for OpenFlow applications to consume network resources, it is vital to set a threshold of the number of flow rules that an OF app can install. If controllers always accept a new flow configuration without any limitation of access rate, the surge of calling requests from malicious applications can create the downgrade of controller performance. Even worse, it can lead to crashes and failure of entire network operation \cite{c30}. Thus, we need to keep network applications under observation and limit their quota of consuming resources based on their priority and granted permission.
\subsubsection{Information disclosure}
Due to the shortage of supporting encryption in the Northbound interface, the data exchanged between SDN controller and applications can be captured and tampered \cite{c25}. Besides that, a malicious application can directly perform reconnaissance attacks to gather network information through API abusing since there is no protection system located at the Northbound interface. Consequently, attackers can use these sensitive resources such as switch ports, switch links and flow rules to prepare sophisticated attacks like DDoS, topology poisoning, etc. Hence, it is critical to build a repacking service to control and observe data transferred from an SDN controller to an OpenFlow application. This function requires the transparency of the monitoring scheme with OpenFlow apps that is helpful to protect the SDN controller against malicious applications. Consequently, the SDN controller is hidden from maliciously probing due to the interception of the monitoring system. Additionally, the Northbound API encounters a lack of encryption services for communication between network applications and controllers, which can ultimately result in being susceptible to the Man-in-the-middle attack.
\subsection{Research goals}
It is evident that the whole network operation can be broken by rogue applications through the Northbound interface provided by the controller, according to a previous study \cite{c14}. The malicious or compromised applications can dynamically change the services of other applications or perform harmful intent without any constraint. The reason for this is that the paradigm of SDN allows a third party to deploy Open-Flow applications to consume network resources via the Northbound interface. If the Northbound API is abused by malicious applications, hackers can perform attacks to compromise the SDN controller. The goal of our study is to introduce a novel fine-grained and decentralized access control framework for OpenFlow applications in the Northbound interface of SDN. Six challenges mentioned before are resolved by 3 solution categories as follows.

Firstly, our approach provides a fine-grained control system built with the main idea of the controller-independent principle. It authenticates every communication and entity joining the application-controller channel to prevent data leakage. This is carried out by checking whether each request for accessing API assets owns the right permission or not.

Secondly, it realizes policy-based access control on the Northbound interface. Therein, the administrator is responsible for creating permission lists for accessing API assets grouped by object types in SDN. Then, for each network application, a policy is defined with granting permissions corresponding with the role profile of the application in the network. Otherwise, it denies access if there is no policy created for this application.

Thirdly, we also ensure the legality and the atomicity of installing flow rules by preventing conflicts within network functions. By recording all behaviors of OpenFlow app communicating with the controller such as API abuse, flow conflict creation, the framework automatically conducts a consideration of application trustworthiness for further incidents response and reporting to administrators.

By utilizing blockchain in the three aspects shown above, we also guarantee the availability and robustness of anti-counterfeiting for monitoring schemes. Our security-enhanced solution for SDN Northbound leverages the key characteristics of the blockchain, namely decentralization, immutability, and auditability, to resist the security concerns about ensuring the integrity and tamper resistance of critical information in the SDN-enabled network.
\begin{figure*}[!t]
\centering
\includegraphics[width=0.6\textwidth]{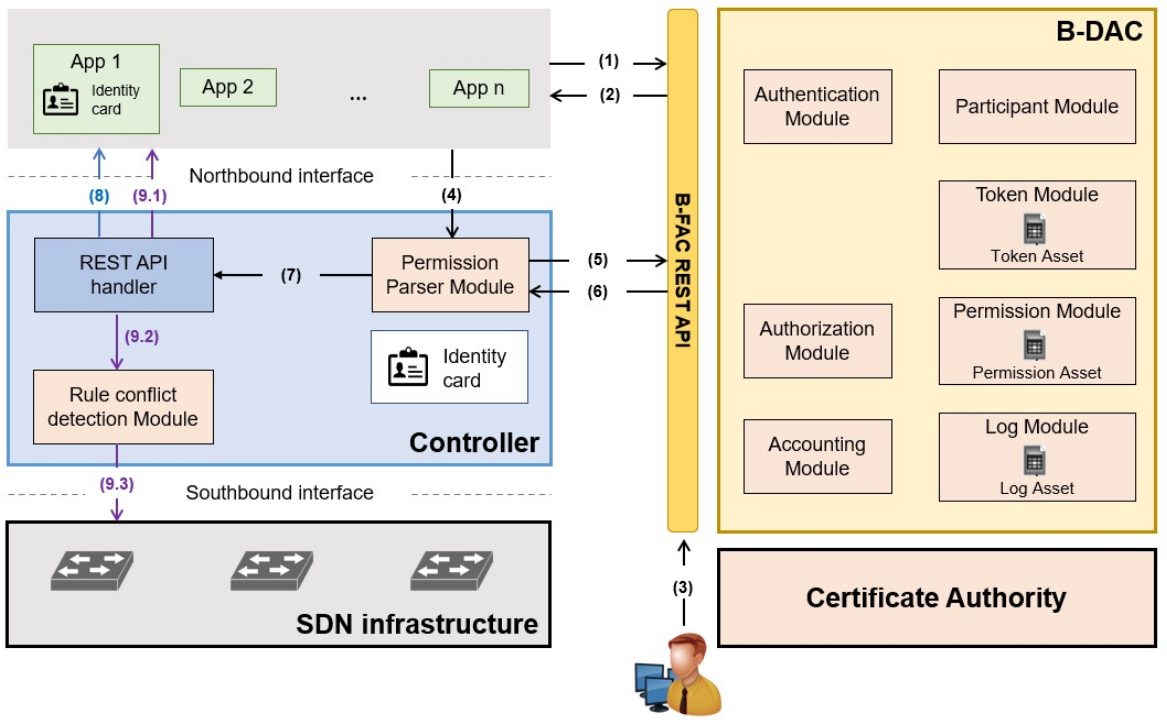}
\caption{Proposed Blockchain-based Access Control for SDN-enabled networks: B-DAC system}
\label{fig2}
\end{figure*}
\section{Architecture Design}
\label{sec4}
To address problems discussed in \textbf{Section~\ref{sec3}}, we introduce an approach of decentralized access control for network applications in SDN using blockchain. According to the highlight features aforementioned in Section \ref{HyperledgerFabric}, we choose Hyperledger Fabric, a permissioned blockchain to deploy an AAA scheme for network security management because it provides the low-cost blockchain approach with privacy-oriented, high-performance and scalable characteristics. The proposed framework in this paper, called B-DAC, is inherited, and extended, from our previous works \cite{c11} \cite{c41}. Our B-DAC consists of multiple modules to guarantee the security of controller-application communication, which is depicted in \textbf{Fig.~\ref{fig2}}. The function of each module is mentioned in detail in the following sections.
\subsection{Entities in B-DAC}
\subsubsection{Participants}
In B-DAC system, blockchain is leveraged to support the operation of the AAA scheme, where SDN components are considered in Blockchain-based terms.

\begin{figure}[!t]
\centering
\includegraphics[width=0.42\textwidth]{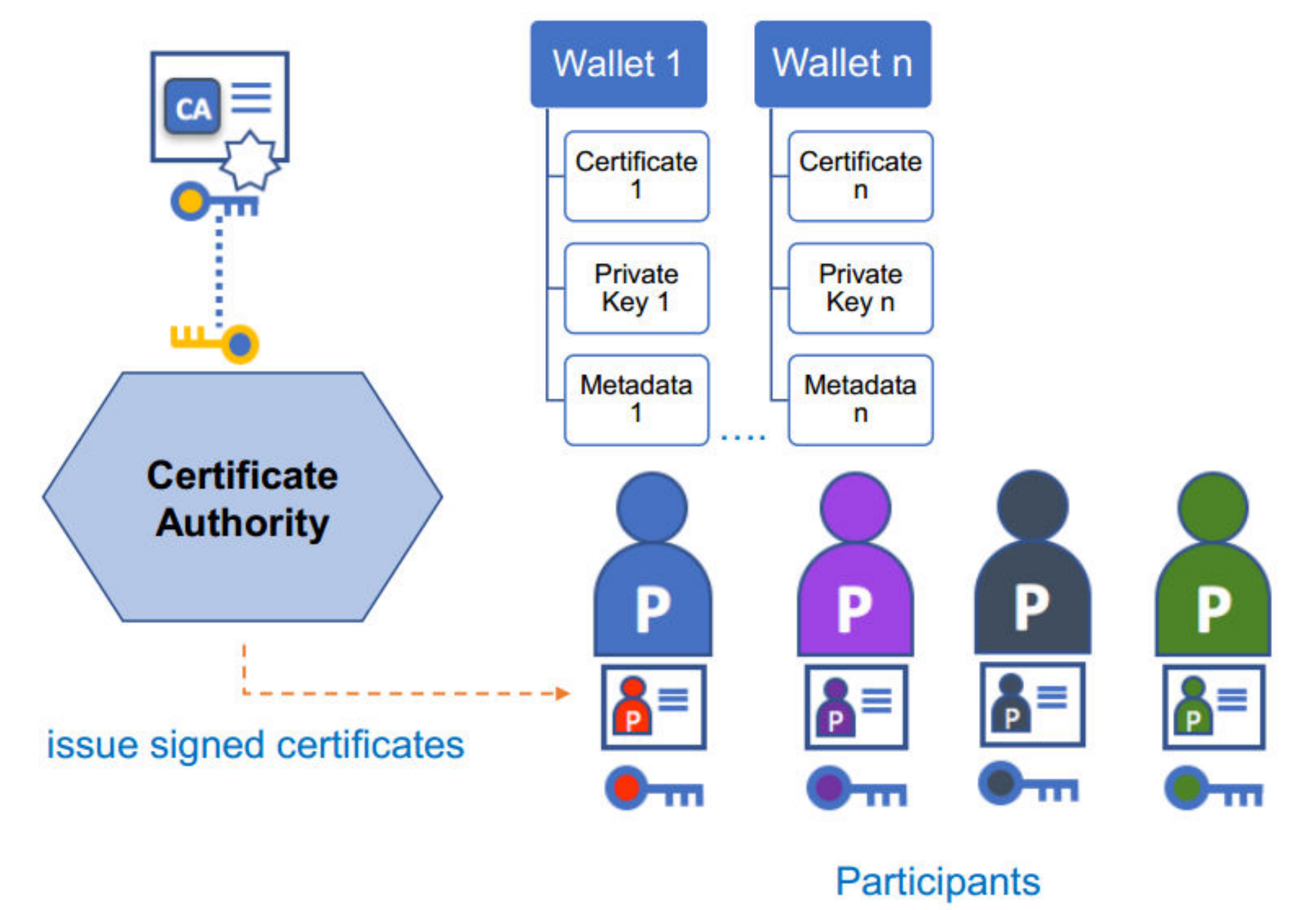}
\caption{The wallet structure of participants with X509 certificates issued by Certificate Authority (CA)}
\label{fig-wallet}
\end{figure}

A participant must have a digital identity which is stored in a wallet to connect to a Hyperledger Fabric. An X509 identity (X509Identity) is created by the network administrator when a participant joins the blockchain network. It is a set of information and credentials of the participant in communication within the system, including a MSP (Membership  Service  Provider) identifier (MSPID in metadata), a private key, and an X.509 certificate issued by CA, as shown in \textbf{Fig.~\ref{fig-wallet}}. The private key in the identity is used for signing the new transactions. Specifically, Elliptic Curve Digital Signature Algorithm (ECDSA) is a digital signature algorithm is used in Hyperledger Fabric, which is more adaptable in the context of IoT \cite{2018_bubleoftrust}. In fact, ECDSA brings many advantages over conventional signature algorithms such as RSA regarding signature times and key sizes, while achieving the same strength.
\[
\textit{Identity =\{X509Certificate, PrivateKey, MSPID\}}
\]

Given a key pair of the private key \textit{K\textsubscript{pr}} and the public key \textit{K\textsubscript{pu}}, where \textit{K\textsubscript{pu}=G*K\textsubscript{pr}} and \textit{G} is a point on the elliptic curve \textit{E}, ECDSA uses a random number \textit{r} to improve the security of the signature. A random number \textit{r} is created before encrypting the plaintext \textit{M}.
The ECDSA scheme of performing private key signature follows the order as in (\ref{eq1}):
\begin{equation} \label{eq1}
\begin{split}
   & \phi = r*G(x, y) \\
 & h = Hash(M) \\
 & s = \frac{(h+K_{pr}*x)}{r} \\
 & \textit{Commit: } Peer Node \leftarrow \{M, (\phi, s)\}
\end{split}
 \end{equation}
 
Afterward, the peer nodes verify the signature \((\phi, s)\) with the public key \textit{K\textsubscript{pu}} as in (\ref{eq2}):
\begin{equation} \label{eq2}
\begin{split}
 & h = Hash (M) \\
 & \psi = \frac{(h*G+x*K_{pu})}{s} \\
 & \textit{Verify: }   \psi  == \phi
\end{split}
\end{equation}

Because the main goal is to protect the communication between the controller and OpenFlow applications, these two components are the main assets in our Blockchain-based system. Each participant is identified by two fields of \emph{id} and \emph{name}. Moreover, participants which are applications have additional fields in their profiles. There is a field called \emph{role} to store their role assigned by the system.

To enable the trust establishment, B-DAC framework evaluates the trustworthiness of applications in a specific network, relying on their behaviors. In our design, along with basic necessary information for application authentication, we also declare an additional field in the profile of an application, called \emph{Trust Index}. It is used to reflect the well-behaved level of a specific application and determine whether the SDN controllers should trust or deny its further actions. The value of \emph{Trust Index} will be decreased in case of overprivileged attempts or conflicts in flow rules installation from OpenFlow applications.

\begin{figure}[!b]
\centering
\includegraphics[width=0.4\textwidth]{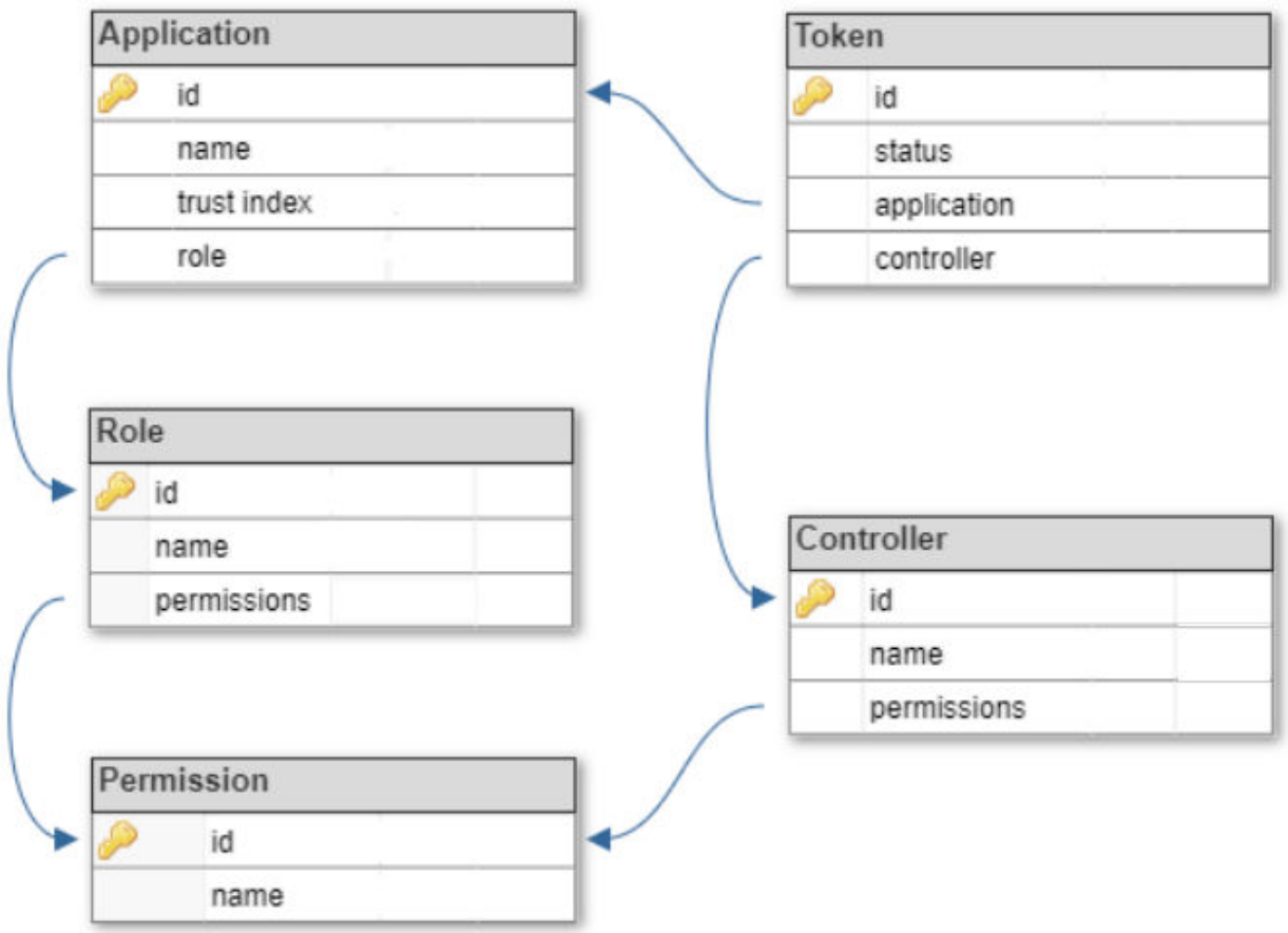}
\caption{The structure of main assets in B-DAC}
\label{fig-app-controller}
\end{figure}

Note that, \emph{Participant module} manages all the participants in the blockchain. The asset structures of OpenFlow applications and SDN controller in B-DAC are depicted in \textbf{Fig.~\ref{fig-app-controller}}.

\subsubsection{Main functional components in B-DAC}
To support the operation of authentication, we use a token to only allow registered applications to interact with the system. These tokens are created and managed by the \emph{Token module} of the system. In addition, the authorization function requires pre-defined permissions to check whether the action of the application is over-privileged and needs to be noticed. Therefore, we design \emph{Permission module} to manage Permission objects. Both token and permissions have their representation as important kinds of blockchain asset, as illustrated in \textbf{Fig.~\ref{fig-app-controller}}.

\begin{figure}[!b]
\centering
\includegraphics[width=0.48\textwidth]{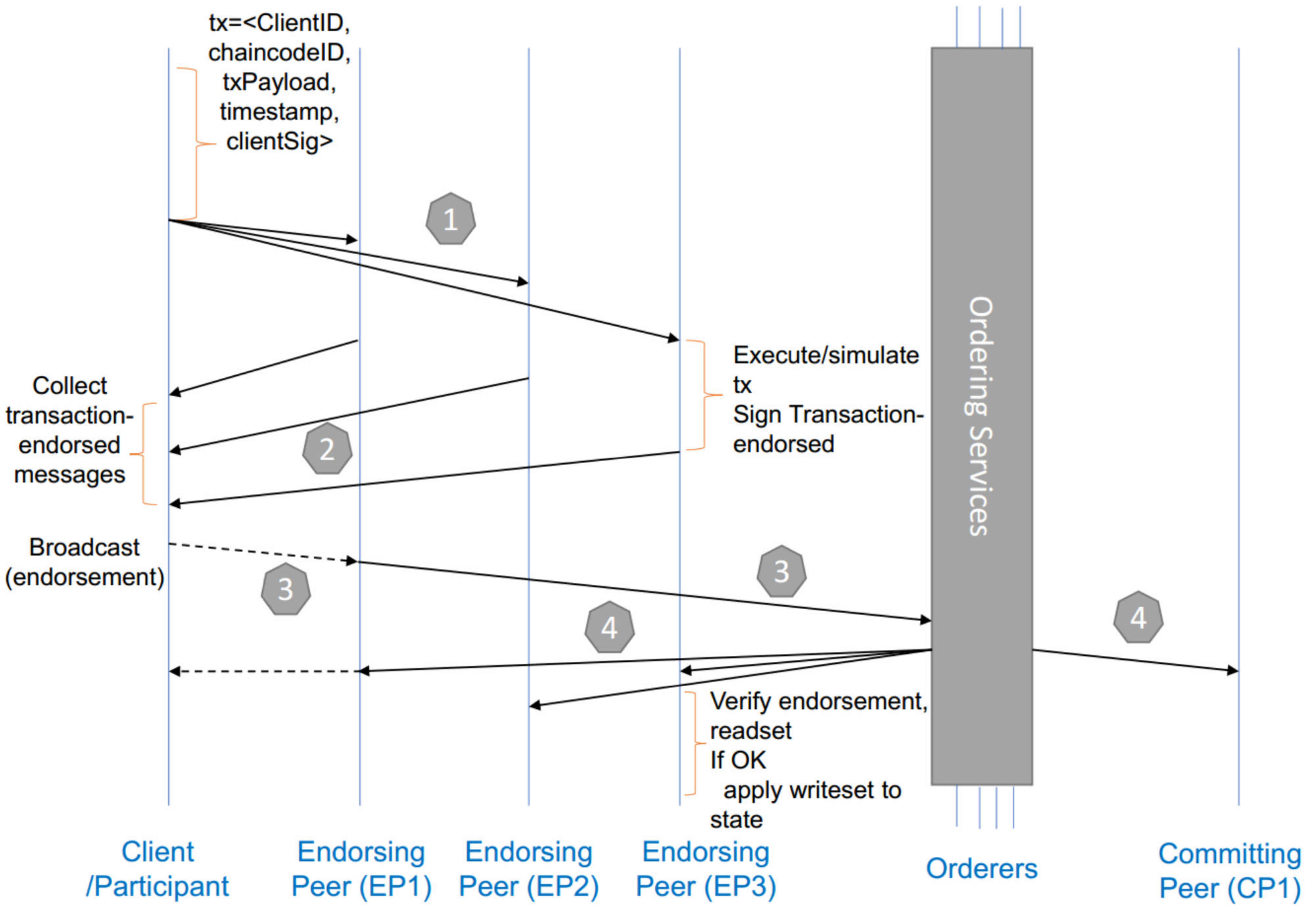}
\caption{The transaction flow in Hyperledger Fabric-based B-DAC system}
\label{fig-transaction-flow}
\end{figure}

Besides, logging is the basic function of accounting, where log entry is obtained and saved in a specific format. Our design has the \emph{Log module} taken this duty. Notably, log entries are considered the main type of asset in the blockchain network. All actions that intend to change the value of assets in the distributed ledger (blockchain) are performed by transactions, as given in  \textbf{Table~\ref{table-transaction-structure}}.  Meanwhile, \textbf{Fig.~\ref{fig-transaction-flow}} presents the flow of a transaction from initiating state at clients/participants to committing state in the blockchain.

\begin{table}[t]
\caption{The structure of transactions in B-DAC}
\centering
\begin{tabular}{|l|l|l|}
\hline
\textbf{\textit{Transaction Type}} & \textbf{\textit{Participant}} & {\textbf{\textit{Transaction Payload}}} \\ \hline
addApplication   & Admin       &id, name, trust-index, role-id  \\ \hline
updateApplication   & Admin       &app-id, name \\ \hline
updateAppRole   &Admin  &app-id, role-id    \\ \hline
updateAppTrustIndex &\makecell[l]{Admin,\\ Controller}  &app-id, trust-index \\ \hline
removeApplication   &Admin  &app-id \\ \hline

addController    & Admin       &id, name, permissions   \\ \hline
updateController    &Admin  &controller-id, name, permissions    \\ \hline
removeControleler   &Admin  &controller-id  \\ \hline

createPermission & Admin       &id, name    \\ \hline
removePermission & Admin       &permission-id   \\ \hline

createRole       & Admin       &id, name, permissions   \\ \hline
updateRole       & Admin       &role-id, name, permissions  \\ \hline

requestAppToken  & Application &controller-id   \\ \hline
issueToken       & Admin       &token-id    \\ \hline
expireToken      & \makecell[l]{Admin, \\Controller}       &token-id    \\ \hline
addLogEntry      & Controller  &\makecell[l]{resource-url, data, token-id, \\ http-method, permission-id, \\app-id, controller-id, \\action, message}\\ \hline
\end{tabular}
\label{table-transaction-structure}
\end{table}

\subsubsection{RESTful API for AAA scheme}
To enhance the flexibility and independence of our system on the operation of any specific controller, we implement the AAA scheme outside the controller. More specifics, B-DAC provides RESTful APIs for controllers as well as OpenFlow applications to interact with it. Participants (i.e., controller and applications) can send HTTP requests to this REST API to query or update information in the network system, respectively. Moreover, to ensure the security of exchanged information, HTTPS is enabled for this communication.

\subsection{Policy definition}
\subsubsection{Request-based Permission principle}
A controller provides APIs for accessing its resources at the Northbound interface of SDN. Each API is distinguished by a URI and an HTTP method. The most well-known methods for HTTP are POST, GET, PUT and DELETE that correspond to operations of creating, reading, updating, and deleting, respectively. Thereby, URI is used to locate the network resource. A pair of URI and HTTP method offers an action on a network resource of the network. In fact, we define a set of permissions corresponding to APIs provided by the controller. Detailed information such as access token, the content type is carried in the header of the HTTP request. For each request from an application to a controller, parameters of URI and HTTP method are fetched and processed by \emph{Permission Parser} in SDN controller to determine required permission for this request.
\subsubsection{Policy definition}
In B-DAC, to protect network resources from unauthorized requests, a policy is designed to determine whether to approve or reject a REST API request stemming from an OpenFlow application. In detail, we divide the policy into two categories, including Role Policy and Trust Policy. 

First, the policies in the Role group are used to manage the permitted behaviors of an application on specific resources. They determine which REST API requests that one application with an assigned role can perform. It means that if an application is designated to ascertain a role group, it can send commands and receive response information from the controller. In addition, each application role has a priority to control the WRITE actions on its data from other applications. Therein, the application with lower priority cannot update or delete data in the network written by the higher one. To achieve this goal, a role-based access control (RBAC) model is chosen to assign permissions to applications. It is a common approach that helps to control rights more effectively based on each role of the authorized subject. Access can and should be granted on a need-to-know basis. 

With hundreds or thousands of applications, security is more easily maintained by restricting unnecessary access to sensitive information based on each application’s established role within the network. Consequently, administrators can promptly update permissions for multiple applications by changing permissions of a specific role. To start with, administrators define permissions based on APIs provided by the controller. After that, they create roles in which pertinent permissions are inserted (\textit{createRole} transaction). Eventually, an appropriate role is assigned to an OF application for consuming network resources. In the profile of each application, there is a role property, called \textit{role-id} regarding its functions. To grant a set of permissions for a specific application, administrators create an \textit{addApplication} transaction to add the role with \textit{role-id} to their profile. Also, administrators can utilize \textit{updateAppRole} transaction in B-DAC to change the role of an application.

Meanwhile, the Trust Policy triggers the event of revocation of active permission in the application. In fact, each permission is grouped by the type of resources. By default, each group has a minimum value of trustworthiness to allow an application to utilize. This value can be changed for the different levels of security by the administrator. When the trustworthiness in the application profile is below the minimum threshold of the relevant resource object, all issued permissions regarding this object are disabled on this application. To have a comprehensive understanding, a sample of Trust Policy is depicted in \textbf{TABLE~\ref{tab1}}. Note that, each threshold shown in the table is not arranged in any specific order.
\begin{table}[t]
\caption{A sample list of available resources and Trust threshold of corresponding permission}
\begin{center}
\begin{tabular}{|c|c|c|}
\hline
\textbf{\textit{Resource object}}& \textbf{\textit{Default threshold of Trust}}& \textbf{\textit{Permission ID}} \\
\hline
host& THRE\_1 & p1,   \\
switch& THRE\_2 & p2, p3   \\
link& THRE\_3 & p4, p5   \\
port& THRE\_4 & p6   \\
flowmod& THRE\_5 & p7   \\
group& THRE\_6 & p8, p9  \\
vlan& THRE\_7 & p10, p11  \\
statistics& THRE\_8 & p12, p13, p15 \\
application& THRE\_9 & p14\\
controller& THRE\_10 & p16, etc. \\
\hline
\end{tabular}
\label{tab1}
\end{center}
\end{table}

As to policy generation, the network administrator is first offered a full description of API requests according to critical resources. Subsequently, a list of permissions is created for all requests to control their resources. After that, relying on the security requirements, the network administrator defines ACL (access control list) to enforce the fine-grained security policy on the requests. Particularly, there are ACL definitions limiting participants on accessing B-DAC assets. For instance, all applications are prohibited to change their permissions or insert a new role in the network system without the administrator’s acceptance. \emph{app\_X} with role \emph{P} can query the information about \emph{app\_Z} but cannot create a link. Whereas \emph{app\_Y} with role \emph{Q} can enable the firewall in the specific network, etc. When an application joins the network, the administrator only needs to designate this application with the corresponding role (new or existing). Then, this application must comply with these policies, otherwise, it is rejected or blocked for further requests. \textbf{TABLE~\ref{tab2}} shows the sample ACL list which is defined by the administrator. It is notable that there is always a DENY ALL rule at the end of the ACL.

\begin{table}[!t]
\caption{ACL samples for policy definition in B-DAC}
\begin{center}
\begin{tabular}{|c|c|}
\hline
\textbf{Participant} & Network Administrator \\
\textbf{Operation}&  ALL  \\
\textbf{Resource}&  ALL   \\
\textbf{Condition}&  None  \\
\textbf{Action}&  ALLOW \\
\textbf{Description}&  \makecell{Allow Network Administrator to perform all \\ operations (CREATE, READ, UPDATE, DELETE) \\ on all resources.} \\
\hline
\hline
\textbf{Participant} & Application (p) \\
\textbf{Operation}&  READ  \\
\textbf{Resource}&  Application (r)   \\
\textbf{Condition}&  p.id == r.id  \\
\textbf{Action}&  ALLOW \\
\textbf{Description}&  \makecell{The application is only allowed to read its \\ information by itself.} \\
\hline
\hline
\textbf{Participant} & Application \\
\textbf{Operation}&  CREATE  \\
\textbf{Resource}&  Token   \\
\textbf{Condition}&  None  \\
\textbf{Action}&  ALLOW \\
\textbf{Description}&  \makecell{Allows an application to create new tokens.} \\
\hline
\hline
\textbf{Participant} & Application (p) \\
\textbf{Operation}&  READ  \\
\textbf{Resource}&  Token (r)   \\
\textbf{Condition}&  r.application.id == p.id  \\
\textbf{Action}&  ALLOW \\
\textbf{Description}&  \makecell{The application is only allowed to read the tokens \\ created by it.} \\
\hline
\hline
\textbf{Participant} & Controller (p) \\
\textbf{Operation}&  READ  \\
\textbf{Resource}& Controller (r) \\
\textbf{Condition}&  r.id == p.id  \\
\textbf{Action}&  READ \\
\textbf{Description}&  \makecell{The controller is only allowed to read its \\ information by itself.} \\
\hline
\hline
\textbf{Participant} & Controller \\
\textbf{Operation}&  CREATE  \\
\textbf{Resource}&  verifyRequest   \\
\textbf{Condition}&  None  \\
\textbf{Action}&  ALLOW \\
\textbf{Description}&  \makecell{Allow Permission Parser of the controller to send \\ requests to verify permissions to the system.} \\
\hline
\end{tabular}
\label{tab2}
\end{center}
\end{table}
\subsection{Detailed scheme of AAA}

\subsubsection{Authentication}
The authentication relies on a module called \emph{Authentication module} on the B-DAC system. The responsibility of this module is to verify the identity of applications or controllers in the system using their wallets. Our system uses the certificate-based authentication scheme, with certificates consisting of identity information instead of traditional username-password pairs. Therefore, a \emph{Certificate Authority} (CA) is required to support this function.

In the context of B-DAC model, there are two participants that need to be authenticated, namely applications and controllers when utilizing SDN resources. It is helpful to resolve the spoofing problem of SDN applications and controllers.
\paragraph{Authenticate to B-DAC's REST API}
\label{sec5.4.1.1}
Because all interaction with B-DAC system is via its REST API, every participant first needs to authenticate to this API before any further actions. We use JSON Web Token (JWT) \cite{c42} for this purpose. Based on the information of requesting user and secret key, B-DAC returns a corresponding access token to authenticate with the B-DAC REST API.

However, after successfully accessing the REST API, participants are still required to upload their identity cards (according to wallets) to B-DAC system. This card is then used to authenticate with the Blockchain and sign transactions, as depicted in \textbf{Fig.~\ref{fig-wallet}} above. The wallet address enables the application authentication: an unknown address indicates that the application is not registered to interact with B-DAC and the controller.

In short, all REST API requests in B-DAC are checked to confirm a valid participant by the JWT access token and identity card. Otherwise, it rejects the request.
\paragraph{Authenticate OpenFlow application}
\label{sec5.4.1.2}

Once an application wants to connect to a controller, an additional token provided by the B-DAC system is required. This token is a representative of access requests from a specific application to a controller; hence, it contains information about these two objects. This token can have one of three statuses, which are NEW, ISSUED and EXPIRED. Only ISSUED tokens can be used in the AAA scheme, others require further actions from the network administrator to be useable.

To get its token, the application must specify in detail which controller that it intends to connect to and send a request to B-DAC REST API. Specifically, such input parameters are structured in a transaction proposal (\textit{requestAppToken} transaction in \textbf{TABLE~\ref{table-transaction-structure}}) to be processed by peer nodes in the blockchain network. The participant's cryptography credential (private key) is used to produce a unique signature for this transaction proposal.

By comparing the information in the incoming request and the endorsement policy in the blockchain network, B-DAC promptly decides on releasing a token or not. A token request is valid if: (1) it is sent from an application, (2) it has the information about the requested controller. If everything is valid, B-DAC creates a NEW-state token including the information about the corresponding application and controller as well as its creation time. Then, this token is sent back to the application as a response. However, the NEW token needs to wait for further consideration of the administrator to be released as ISSUED one and becomes ready to use, then. Actions of network administrators like changing a token status are performed by Hyperledger Fabric's chaincode execution with their signatures.

More importantly, the application must integrate this token to every later request to the controller as proof of its authenticated state. The main steps of the authentication scheme are illustrated in \textbf{Algorithm~\ref{al1}}.
\begin{megaalgorithm}[!t]
\begin{algorithmic}[1]
    \Require 
    \Statex \emph{controller}: SDN controller,
    \Statex \emph{app}: application,
    \Statex \emph{bdac}: B-DAC database
    \Ensure ACCEPT/DENY
    \State approved $\leftarrow$ false
    \State /* first check REST API authentication of B-DAC
    \State with JWT access token and identity card */
    \If{app.JWTtoken \&\& app.identity \textbf{exists} in B-DAC}
        \State token $\leftarrow$ bdac.getToken(app, controller)
        \State /* then checking the app-controller authentication */
        \If{token \textbf{exists}}
            \Switch{token.status }
                \Case{NEW}
                \EndCase
                \Case{EXPIRED}
                  \Assert{approved $\leftarrow$ false}
                \EndCase
                \Case{ISSUED}
                  \Assert{approved $\leftarrow$ true}
                \EndCase
            \EndSwitch
        \EndIf
    \EndIf
    \If{approved == true}
        \State \textbf{return} ACCEPT
    \EndIf
    \State \textbf{return} DENY
\end{algorithmic}
\caption{Performing API Authentication scheme} \label{al1}
\end{megaalgorithm}

Created tokens are saved in \emph{Token Asset} of the Blockchain network and available for later access. All token-related operations of the system are processed by the \emph{Token module}.
\subsubsection{Authorization}
\label{sec5.4.2}
The authorization operations are applied to check whether an application has enough permission to request a specific SDN resource or perform an action on the network. Upon obtaining proper tokens, an application can interact with the controller and send requests as normal. Later, the authorization and accounting processes are indeed  communications between only the controller and B-DAC system using tokens, which are transparent to applications. The overall process of fetching permissions to authorize application requests is shown in \textbf{Algorithm~\ref{al2}}.

\begin{megaalgorithm}[!t]
\begin{algorithmic}[1]
    \Require 
    \Statex \emph{bdac}: B-DAC database
    \Statex \emph{permission}: parsed permission,
    \Statex \emph{app}: OpenFlow application,
    \Statex \emph{app\_permissions}: permissions of application in B-DAC
    \Ensure ACCEPT/DENY
    \State /* B-DAC received a verifying request from controller that contains Input */
    
    \If{ bdac.getParticipant().type is CONTROLLER  \&\&  
    \State bdac.getParticipant() == app.token.getController() \&\&
    \State app.token.status == ISSUED \&\&
    \State permission is in app\_permissions \&\&
    \State app.trust\_index $\geq$ permission.trustThreshold}
        \State \textbf{return} ACCEPT
    \Else
        \State app.trust\_index decrease by 1
    \EndIf
\end{algorithmic}
\caption{Performing Authorization process} \label{al2}
\end{megaalgorithm}

 B-DAC has the \emph{Permission Parser} and \emph{Authorization module} to provide permission-related functions. We develop Permission Parser located on the SDN controller to interpret incoming requests for further verification in B-DAC system. The built-in Permission Parser module is used to parse permissions from the URL in the request. Specifically, the parsing process tries to map the API in the incoming request to corresponding permissions. Then, the request information consisting of the URL, data, HTTP method, token and parsed permissions is sent from the controller to B-DAC.

Data sent from \emph{Permission Parser} then is processed by the \emph{Authorization module} in B-DAC. The permission granted to the application (stored in B-DAC database) is compared to the parsed information from the requested API. So, a request must satisfy all following criteria to be considered as a valid one: (1) it is sent from a valid controller, (2) the source controller must be the same as the controller mentioned in the access token, (3) the application in access token must be valid and exists, (4) the parsed permissions must be a subset of the granted permissions of the application, (5) the Trust Index of requested application is higher than the trust threshold of permission. The description of \emph{Trust Index} is discussed in more details in \textbf{Section~\ref{sec5.6}} below. Consequently, an ACCEPT result is responded to the controller when all the above criteria are met, otherwise, a DENY one is returned.

When it comes to the trustworthiness of application, we inherit the trust-based management scheme from \cite{c11} to consider the trust of B-DAC system on an application. Based on the permission checking process, any over-privileged action of application is punished by decreasing its \emph{Trust Index} through\textit{ updateAppTrustIndex} transaction. This means that applications with well-behaviors remain their \emph{Trust Index} in high value from the B-DAC. When this field is lower than a specific safe value, the application is automatically considered as an untrusted one. Depending on the security policy defined by the administrator, untrusted applications can be blocked or disabled. 
\subsubsection{Accounting}
All operations of B-DAC system, as well as the requests sent from applications to controllers, are recorded in log entries and transactions in the Blockchain network (\textit{addLogEntry} transaction). Each log entry contains the essential information about the request from an application, which is managed and reviewed via \emph{Log module}. The structure of log entry is illustrated in \textbf{TABLE~\ref{tab3}}. This is a standard form for each request of verifying permission. Obviously, it is crucial to guarantee the integrity of audit logs for investigation in case of unexpected network failures. With the aforementioned features, blockchain is the relevant approach for this problem.

\begin{table}[!t]
\caption{The structure of Log entry for the request of permission verification}
\begin{center}
\begin{tabular}{|c|c|}
\hline
\textbf{\textit{Log field}}& \textbf{\textit{Description}} \\
\hline
id & Identification of entry log\\
created\_time &  Date time of entry creation\\
url &   URL of request sent by application\\
data &  Payload of request \\
token\_id & Token\\
http\_method & HTTP method used in the request\\
permission\_id &  Permission of application\\
application\_id &   Identification of application\\
controller\_id &  Identification of controller  \\
action & DENY or ACCEPT result of request\\
message & Notification of system\\
\hline
\end{tabular}
\label{tab3}
\end{center}
\end{table}

\subsection{Flow rule conflict detector}
In addition to application authentication and authorization tasks, our system also aims to prevent compromised applications from misusing flow rules to affect the network. Moreover, it can control many conflicting rules in the OpenFlow switches to avoid the issue of flow redundancy. A module, named \emph{Flow rule conflict detector module}, is built in the controller to meet these requirements.

In fact, not every request can cause a flow conflict problem, this issue only occurs when there are flow rule modifications in flow tables. Hence, only requests from an application that lead to this change need to be considered by our module. Otherwise, we let other requests be processed as usual without any doubt about malicious flow rules. To distinguish these types of requests, we use the assumption that applications attempt to use supported RESTful APIs from the controller to perform their job and interact with the SDN network. Then, our conflict detection module bases on APIs and their functions to figure out which request we should take care of. In other words, only RESTful APIs used to modify flow tables and their requests are analyzed in this module. For example, we have considered ACL and Firewall APIs in Floodlight controller in this work. These APIs have the flow rule defined in their formats, sent in the request body. Our module attempts to intercept and analyze these rules to look for any conflicts.

\begin{figure}[!t]
\centering
\includegraphics[width=0.28\textwidth]{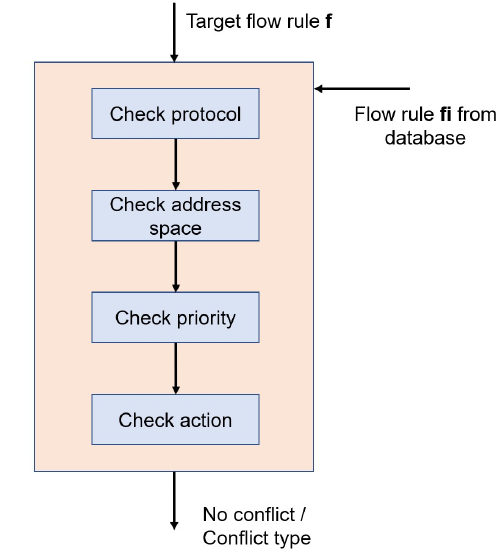}
\caption{Scheme of flow rule conflict detection on target flow $f$ and $f_i$ saved in network flow rules database.}
\label{fig3}
\end{figure}

Once capturing a request to such APIs, this module extracts multiple fields to reform a flow rule. Later, this rule is checked against the pre-defined conflict policy with two steps:
\begin{enumerate}
    \item Verify the validation of flow rule: this step extracts all parameters and their values from the flow rule. Then, these parameters are checked if they are supported as well as their value does not violate the restriction like range, maximum value, etc. The lists and ranges of parameters may be different from an API to others, which have their methods of defining their rules before converting to the actual flow rule entries.
    \item Detect and classify flow rule confliction: this step inherits the method of detecting flow rule confliction from \cite{c40}. The detector compares a target flow rule to a saved conflict-free flow rules database to figure out if any confliction exists. Not all but only 4 fields are considered, which are \emph{protocol}, \emph{dst\_ip}, \emph{priority} and \emph{action} in their order of analyzing, shown in \textbf{Fig.~\ref{fig3}}. With each pair of rules, those fields are verified according to the detecting algorithm of \cite{c40}, which output is the corresponding confliction type. This work also defines 5 conflict types of Generalization, Redundancy, Correlation, Shadowing and Overlap. Once a conflict is detected, our system simply denies this flow rule with CONFLICTED signal.
\end{enumerate}

\subsection{Managing the Trust Index of applications in B-DAC system}
\label{sec5.6}

\begin{megaalgorithm}[!t]
\begin{algorithmic}
    \Require 
    \Statex \emph{request}: a REST API request,
    \Statex \emph{trust}: Trust Index of application,
    \Statex \emph{profile}: application permission profile,
    \Statex \emph{policy}: permission policy on resource object,
    \Statex \emph{flows}: flow rules set
    \Ensure ACCEPT/DENY
    \State /* B-DAC received a verifying request from controller that contains Input */
    
    \If{request \textbf{unmatches} profile $\parallel$ request \textbf{conflicts} flows}
        \State trust $\leftarrow$ trust -1
        \State object $\leftarrow$ request.toResourceObject
    \EndIf
    \If{trust $<$ policy.trustThreshold(object)}
        \State profile.permission(request) $\leftarrow$ BLOCK
        \State \textbf{return} DENY
    \EndIf
    \State \textbf{return} ACCEPT
\end{algorithmic}
\caption{Dynamically monitoring application’s behavior with Trust Index} \label{al3}
\end{megaalgorithm}

In compliance with the security policy, each permission in our system has the corresponding threshold of Trust Index to allow an application to utilize. So, one application can only activate any permission in its profile when the \emph{Trust Index} is higher than a pre-defined threshold. This allows our system to enforce the policy of permission adjustment at the run-time. Accordingly, B-DAC can automatically downgrade the application permission scope based on its behaviors. The overall process of \emph{Trust Index}  management for dynamically monitoring application is formalized in \textbf{Algorithm~\ref{al3}}. For instance, an application \emph{app1} by default has the value of 100 points \emph{Trust Index} in our design. B-DAC then grants three different permissions to this application, called \emph{p1}, \emph{p2} and \emph{p3} with the \emph{Trust Index} threshold of 80, 75 and 70, respectively. Whenever \emph{app1} sends an over-privileged or unauthorized request, B-DAC decreases the value of the \emph{Trust Index} by one point. Specifically, B-DAC automatically executes the Hyperledger chaincode to make the \textit{updateAppTrustIndex} transaction for downgrading the trustworthiness of the OpenFlow application.

Suppose that the Trust Index of \emph{app1} drops under the threshold of \emph{p1}, i.e. 80 points, our mechanism triggers the partial suspension of granted actions by disabling this permission of the application. Meanwhile, two other permissions \emph{p2} and \emph{p3} still allow \emph{app1} to send further requests not relating to \emph{p1}. Thus, if an application's request violates one or more policies in B-DAC, it may be rejected to be processed by the controller. In this case, there are only administrators who have the capability to recover the trust level from the SDN controller. This action is to help this application properly perform network functions again. All logs are recorded with a timestamp for audit trail in the future.
\subsection{Cache-based performance enhancement in B-DAC system}
\label{sec5.7}
Most of the reading requests, appended with GET method, are only used to query specific information from the SDN network. These requests seem not to make any change in the configuration or data like writing requests (use other methods such as POST, PUT or DELETE). Normally, reading requests are frequently used in applications without significant impact on the network directly. For instance, typical monitoring applications usually send only most of the requests with reading permission. Therefore, we use a caching mechanism on controllers to temporarily save the latest response of B-DAC, which is ACCEPT or DENY signal, for a specific GET request. For more details, with a new GET request whose signal is not available in the cache, the controller sends a verifying request to B-DAC. Then, a corresponding response is sent back and cached in the controller. In the other case, if the signal is available with ACCEPT, the controller responds with the queried data to the application, otherwise, it rejects the request. After that, for all API requests on the same resource, the controller uses the signal in the cache to send a response to the application. Parallelly, to update the latest signal in the cache, another verification request is dispatched from the controller to B-DAC.

\section{The workflow of B-DAC for processing OpenFlow application requests}\label{sec5}
\subsection{Workflow of processing a request sample}
The workflow of processing a request from an OpenFlow application to SDN controller is illustrated in \textbf{Fig.~\ref{fig2}}, consisting of 9 steps. Note that, this process is taken place upon both application and controller have got authenticated to B-DAC REST API by \emph{JWT access\_token} and their identity cards uploaded.
\begin{enumerate}
    \item The new application sends a request for token issuance to B-DAC, which specifies the controller that wants to connect to.
    \item B-DAC authenticates the application and verifies the validity of the request. The valid request that satisfies all requirements mentioned in \textbf{Section~\ref{sec5.4.1.2}} allows B-DAC to create a token with NEW status. Then, this token is sent back to the application.
    \item Subsequently, the network administrator must conduct a manual task of verifying and assigning the ISSUED status to the token if everything is valid. From now, the ISSUED token can be used for the application operation, otherwise, the application with the NEW token is not allowed to access the SDN controller.
    \item The application dispatches all later requests with the encapsulated token to the SDN controller to consume the network resource.
    \item After receiving a request from an application, \emph{Permission Parser} in the controller extract the URL, data, HTTP method, and token to determine the corresponding permission of that application. Next, such data values are sent in the form of a verification request to the B-DAC system.
    However, to speed up the response time of B-DAC, we utilize the caching strategy for requests which had been processed before. While a new request is forwarded to the B-DAC system for authentication, the cached information helps the SDN controller immediately respond to the request of application. Note that, even the response of a request is available to be used, a verification request is still sent to B-DAC. This is not only for accounting, but also for updating the cached information and the \textit{Trust Index} correspondingly. 
    \item Once receiving a verification request from a controller, B-DAC first checks the credential of this controller, then verifies the validity of the request. When all conditions listed in \textbf{Section~\ref{sec5.4.2}} are met, the request is considered as valid and an ACCEPT response is returned to the controller, otherwise, DENY one. Any request with DENY result will trigger to update the \emph{Trust Index} of the requesting application, according to the mechanism aforementioned in \textbf{Section~\ref{sec5.6}}. Moreover, the \emph{Accounting module} creates a log entry in \emph{Log Asset} for verification requests that it receives.
    \item The ACCEPT response from B-DAC indicates that the controller can proceed the request of an OpenFlow application. Otherwise, it rejects this one.
    \item \emph{Flow rule confliction detector} module is used to avoid executing API requests that cause a potential conflict when installing a new flow rule in the SDN. This module checks the possibility of conflict between the intended rule and saved ones. Whether any conflict exists, our module prevents the created flow rule of the request from being sent to devices in the data plane.
    \item Requests that have passed all assessment of B-DAC’s modules can receive their corresponding requested information regarding flow rules installation or statistics retrieval on the data plane.
\end{enumerate}
\subsection{Security characteristics analysis of B-DAC design}
According to the abovementioned design of B-DAC system, it enhances the security of the Northbound interface in SDN by managing OpenFlow applications. It is useful to restrict malicious or unauthorized requests in consuming network resources provided by the controller. In general, our system ensures the following security characteristics:
\subsubsection{Immutability}
By using blockchain technology to manage data, the database of B-DAC cannot be altered or deleted illegally. All transactions containing sensitive data must be validated by all verification nodes before they can be added to the block. Once this block is saved into the Blockchain network, no further modification or removal can be performed.
\subsubsection{Decentralization}
The blockchain network is maintained by a group of nodes rather than a single machine. All these nodes store a replication of the same database, called the digital ledger. Any new coming node is required to clone and store this ledger to be able to join the Blockchain network. A new block added to the ledger is broadcasted to the whole network so that all nodes can record it in their storage. Therefore, the removal of any node does not affect the database stored in the remaining network. This makes the system become decentralized and more available.
\subsubsection{Authentication}
All participants, including SDN controllers and applications, must be authenticated when they want to be involved in B-DAC system. Each of them is offered with an identity card consisting of a private key for authenticating to the B-DAC system and signing transactions. So, only legitimate participants can join and consume network resources.
\subsubsection{Authorization}
Each application has its own granted permissions for accessing controllers. B-DAC offers each application with a corresponding token which contains its privileges for communication with a specific controller. Every request sent from the application to the controller must be integrated with this token. Then, B-DAC verifies the permission of the application to confirm whether the controller should respond or not.
\subsubsection{Accounting}
Since the accounting scheme should consider overcoming the lack of integrity in captured artifacts, there is a need of providing a strong audit trail. So, each activity belonging to third-party apps is stored along with its hash value and timestamp in the form of log entry in B-DAC. The network administrator and participants can access this information to use in authorization control, security analysis or digital forensics. Due to being saved in the Blockchain network, this information is impossible to be removed or modified to ensure the reliability of our system. 
\subsubsection{Flow rules conflicts prevention}
In addition to managing the operation of OpenFlow applications, our design intends to prevent any flow conflicts. Our module analyzes upcoming flow rules insertion requests with installed flows in the data plane to detect rule conflicts. This feature brings the capability of preventing numerous undesired flow rules to drain network resources. Moreover, a malicious application cannot insert any malformed rule to ensure the legality and atomicity of creating flow rules.
\subsubsection{Fine-grained control}
All registered applications are laid under a strict observation by B-DAC. When one application tries to consume the network resources, B-DAC checks the pre-defined behaviors of the application via request's parameters to determine that this application can perform in the network. The application management system does not allow an application to obtain any resources without any limitation. When flow rules are created or updated, each application has a different range of priority in flow tables. Additionally, each application has distinguished permissions corresponding to the roles assigned by the administrator. This security policy enforces the functional hierarchy of applications in the network.
\subsubsection{Trust level management}
By behavior-oriented control, B-DAC framework automatically responds to a rogue OpenFlow application basing on a blockchain-based profile when attackers use compromised applications to scan or probe network resources. More specifically, the trustworthiness of an application can be adjusted dynamically according to their actions with controllers. Taking a compromised SDN application as an indication, its behaviors may harm the network system and should be blocked automatically after several illegal accesses. If the value of Trust Index in the application profile reduces notably, some issued permissions are suspended or revoked to avoid network disruption.
\section{Implementation and Evaluation}
\label{sec6}
We implement the design of B-DAC with the following steps: deploying a Blockchain network and chaincode (smart contract) programming for B-DAC, developing plug-in modules in controller including \emph{Permission Parser} and \emph{Flow rule conflict detector}, for SDN network.
\subsection{Building Blockchain network and deploying B-DAC scheme}
Our system uses Hyperledger Fabric (version 1.4) \cite{c43} as the Blockchain platform for our implementation and evaluation, since it provides the low-cost blockchain approach with privacy-oriented, faster and scalable characteristics. Hyperledger Fabric is a permissioned blockchain allowing to control user access such as network joining, transaction adding or reading, via verifying permission. Moreover, this framework uses the CFT (Crash Fault Tolerant) as a consensus algorithm to guarantee that the system can still correctly reach consensus if nodes fail. Note that, CFT algorithm is cost-free, which becomes its advantage over other counterparts like PoW (Proof of Work) or PoS (Proof of Stake) \cite{c44}. Though BFT (Byzantine Fault Tolerance) is resilient against systems containing malicious actors, it is more complex and expensive than CFT. The CFT consensus algorithm is used to order services implemented with Kafka \cite{c45} and Zookeeper \cite{c46}.

The Blockchain network of B-DAC is deployed on a hardware configuration of 8 cores CPU and 32GB of RAM. This network consists of 17 container-based nodes which are 1 certificate authority, 3 orderers, 3 peers, 3 nodes as the database of peers, 3 nodes for deploying Zookeeper and 4 nodes for Kafka.

Concerning implementation, the enterprise can deploy the blockchain platform relying on the pre-defined architectural design such as the number of peer nodes, orderer nodes, etc. It is also true for a group of enterprises. In the context of B-DAC, the enterprise is the organization that operates the SDN-enabled network and needs to set up the AAA scheme in the blockchain infrastructure. They become the proprietors (stakeholders) of the blockchain network who can validate transactions. After a blockchain network is launched for the first time, its owners (stakeholders) can invite more business partners to co-own the blockchain network. This is done by assigning them validating nodes. Note that, all existing owners need to approve this joining request for any new stakeholder. Then, the new ones can validate the transaction originating from in their business operation or management. Clearly, some different permissioned blockchains under the umbrella of Hyperledger like Hyperledger Fabric are trying to provide interoperability solutions to ensure maximum scalability and adaptability for encouraging intercommunication among different organizations in a specific field, e.g, networking management services in this paper.

After building the blockchain platform with Hyperledger Fabric, we use Hyperledger Composer \cite{hyperledgercomposer} to define the assets and participants of B-DAC. Hyperledger Composer is an extensive, open development toolset and framework which simplifies the process of creating smart contracts (chaincodes) on top of Hyperledger Fabric. Specifically, we write chaincodes corresponding to the payload declaration of transactions for participants or administrators to create or alter these assets in the distributed ledger of Hyperledger Fabric. Also, each participant is granted permissions (CREATE, READ, UPDATE, DELETE) to manipulate on the property of assets by rules in Access Control List (ACL). Subsequently, administrators create a digital identity for each participant in Hyperledger Fabric and save it in a wallet. It is used for participants to make blockchain transactions later.

\subsection{SDN network design}
The controller of our SDN network is a Floodlight controller \cite{floodlight} run on a computer with 4 cores CPU and 4GB RAM. Note that, in our experiment, we implement some network applications by sending requests to the SDN controller to consume network resources through REST APIs of the controller.

\begin{figure}[!t]
\centering
\includegraphics[width=0.35\textwidth]{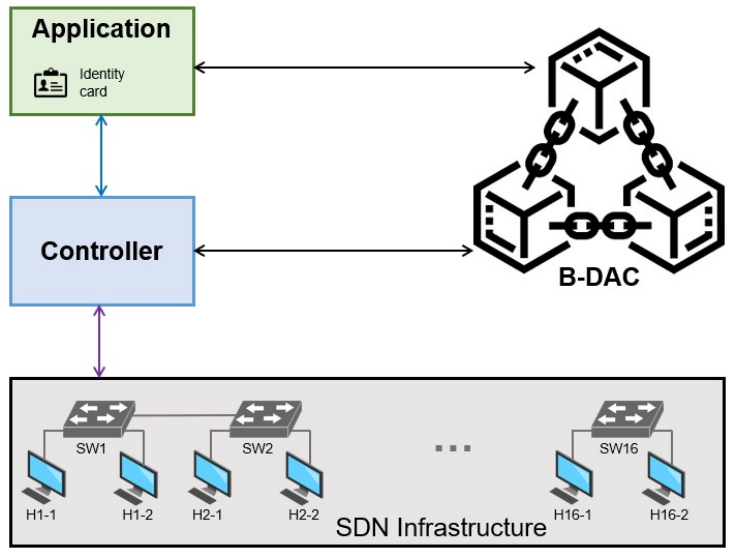}
\caption{Experimental network topology with one controller}
\label{fig4}
\end{figure}

Later, Mininet \cite{c47} is used to simulate an SDN infrastructure layer managed by the above controller. This tool is deployed in another machine having 2 cores CPU and 4GB RAM. This network is a linear-type one consisting of 16 switches and 32 connected hosts, as depicted in \textbf{Fig.~\ref{fig4}}.

In addition, we also develop plug-in modules in Floodlight controller including \emph{Permission Parser} and \emph{Flow rule conflict detector}, for building a B-DAC-enabled SDN controller.
\subsection{Evaluation}
We evaluate the B-DAC on two criteria of performance effectiveness and security assurance for communication between applications and SDN controller at Northbound interface. In our experiment, applications send REST requests to the Floodlight controller through REST API. Following that, B-DAC intercepts these requests to perform access control or enforce security policies at the Northbound. We examine the verification results of those requests in the controller and applications to demonstrate the feature and performance of the proposed system.
\begin{table}[!b]
\caption{Delay time in creating blockchain transaction}
\centering
\begin{center}
\begin{tabular}{|c|c|c|c|}
\hline
\textbf{\textit{\makecell{Number of \\transactions}}}& \textbf{\textit{\makecell{Maximum  \\delay (s)}}} & \textbf{\textit{\makecell{Minimum  \\delay (s)}}} & \textbf{\textit{\makecell{Average \\delay (s)}}} \\
\hline
1000 & 0.95 & 0.31 & 0.39 \\
\hline
\end{tabular}
\label{tab4}
\end{center}
\end{table}

\begin{table}[!b]
\caption{Resource consuming in the blockchain network}
\centering
\begin{center}
\begin{tabular}{|l|c|c|c|c|c|c|}
\hline
\textbf{\textit{\makecell{Node}}}& \textbf{\textit{\makecell{Mem \\(MB)}}} & \textbf{\textit{\makecell{CPU \\(\%)}}} & \textbf{\textit{\makecell{Input \\traffic \\(MB)}}} & \textbf{\textit{\makecell{Out \\traffic \\(MB)}}} & \textbf{\textit{\makecell{Disk \\read \\(MB)}}} & \textbf{\textit{\makecell{Disk \\write \\(MB)}}}\\
\hline
peer0&219.1&5.70&34.2&55.4&0&86.2\\
peer1&281.7&5.72&34.0&55.1&0&86.2\\
peer2&339.8&5.68&34.0&55.1&0&86.3\\
orderer0&19&1.23&20.1&27.5&0&0\\
orderer1&17.2&0.82&11.4&2.5&0&0\\
orderer2&17&0.88&11.7&11.0&0&0\\
kafka0&413.9&5.24&12.5&30.0&0&2.8\\
kafka1&375&2.96&2.7&4.4&0&2.8\\
kafka2&297.3&0.85&0.046&0.072&0&0\\
kafka3&285.8&0.84&0.045&0.071&0&0\\
couchdb0&164.8&40.1&10.9&18.0&2.5&236.2\\
couchdb1&162.9&39.2&10.9&17.9&1.8&236.9\\
couchdb2&162.4&39.2&10.9&17.9&1.4&237\\
zookeeper0&27.5&0.24&0.22&0.13&0&0\\
zookeeper1&26.6&0.29&0.29&0.18&0&0\\
zookeeper2&28&0.35&0.26&0.34&0&0\\
ca&7.5&0.00&0.002&0&0&0\\
\hline
\end{tabular}
\label{tab5}
\end{center}
\end{table}
\subsubsection{Blockchain transaction performance}
Firstly, the performance of Blockchain network is analyzed using Hyperledger Caliper framework \cite{c48}. This blockchain benchmark tool provides the capability of calculating multiple performance metrics such as the number of transactions per second, memory, CPU usage as well as incoming and outgoing traffic, etc. Note that, this tool creates transactions directly according to designed assets in Blockchain network without performing the functionality of B-DAC REST API. \textbf{TABLE~\ref{tab4}} and \textbf{TABLE~\ref{tab5}} show transaction delay and consumed resources of our Blockchain nodes. The results are monitored when our system creates and sends 1000 transactions to the blockchain network. Depending on the number of exchange messages that needs to be processed and confirmed from different parties within the given period, the transaction speed may fluctuate.

Besides, we also consider the response time of our B-DAC system to application requests. The expected time includes authentication, authorization and accounting process when receiving requests from an application. This processing speed may also vary due to blockchain network congestion at the given time when making the transaction. We perform the experiments by sending 1000 requests and computing the average response time, shown in \textbf{TABLE~\ref{tab6}}. Also noted in this table, the caching mechanism mentioned in \textbf{Section~\ref{sec5.7}} is also proved to be effective in reducing the response time approximately 40 times than usual. 
  
To evaluate the performance of B-DAC with other access control schemes for Northbound interface, we compare it to SEAPP approach \cite{c25}, a  secure application management framework based on REST API access control in the SDN Northbound interface. It is independent with SDN controller but prone to the Single Point of Failure.
  
In comparison with SEAPP \cite{c25}, our work has an average latency of 15.6 milliseconds (caching enabled mode) which is nearly 2 times higher than the delay of 7.9 milliseconds in SEAPP. Notably, the SEAPP and B-DAC are both deployed on the same hardware configuration with 32 GB RAM. Although there is an increase in response time stemming from the blockchain process, the response time of B-DAC is acceptable for tackling the Single Point of Failure issue of SEAPP system. The other metrics for evaluating access control scheme, such as CPU/RAM usage of blockchain cluster is much different meaning from a centralized approach like SEAPP; thus we cannot compare these aspects. Nevertheless, we also perform experiments on the scalability of blockchain to measure B-DAC performance below.

\begin{table}[!t]
\caption{Comparison of B-DAC and SEAPP approach in response time to an incoming request }
\begin{center}
\begin{tabular}{|c|c|c|c|}
\hline
\textbf{\textit{\makecell{}}}& \textbf{\textit{\makecell{Maximum (s)}}} & \textbf{\textit{\makecell{Minimum (s)}}} & \textbf{\textit{\makecell{Average (s)}}} \\
\hline
\makecell{B-DAC - No caching}  & 1.03544 & 0.40839 & 0.45088 \\
\makecell{B-DAC - Caching} & 0.37706 & 0.00674 & 0.01566 \\
\makecell{SEAPP approach\cite{c25}}  &$\approx$ 0.009 & $\approx$ 0.004 &  $\approx$ 0.0079 \\
\hline
\end{tabular}
\label{tab6}
\end{center}
\end{table}

\begin{figure}[!t]
\centering
\includegraphics[width=0.48\textwidth]{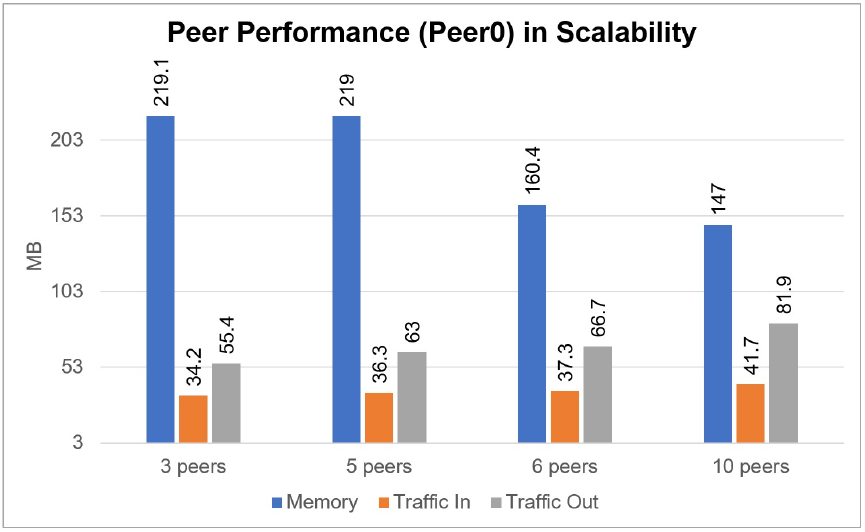}
\caption{Peer performance of B-DAC in scalability}
\label{fig5}
\end{figure}

\begin{figure}[!t]
\centering
\includegraphics[width=0.48\textwidth]{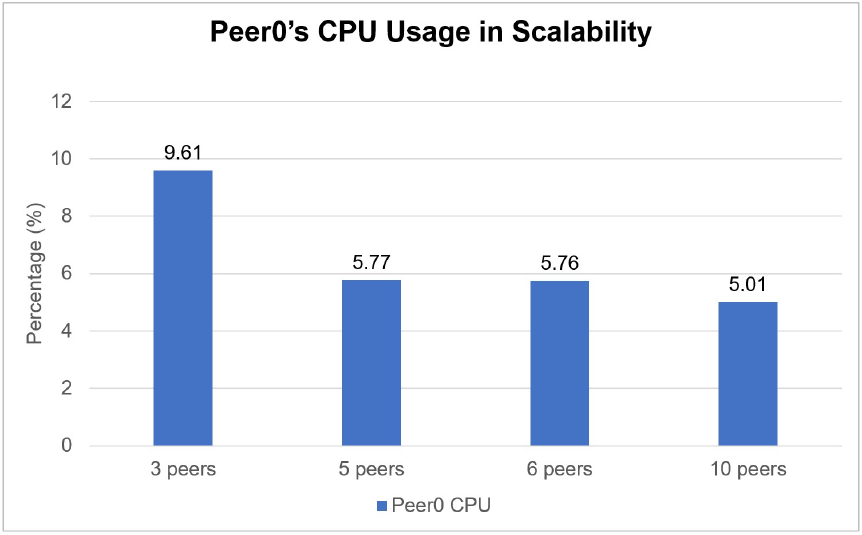}
\caption{CPU usage of B-DAC in scalability}
\label{fig6}
\end{figure}

\begin{figure}[!t]
\centering
\includegraphics[width=0.48\textwidth]{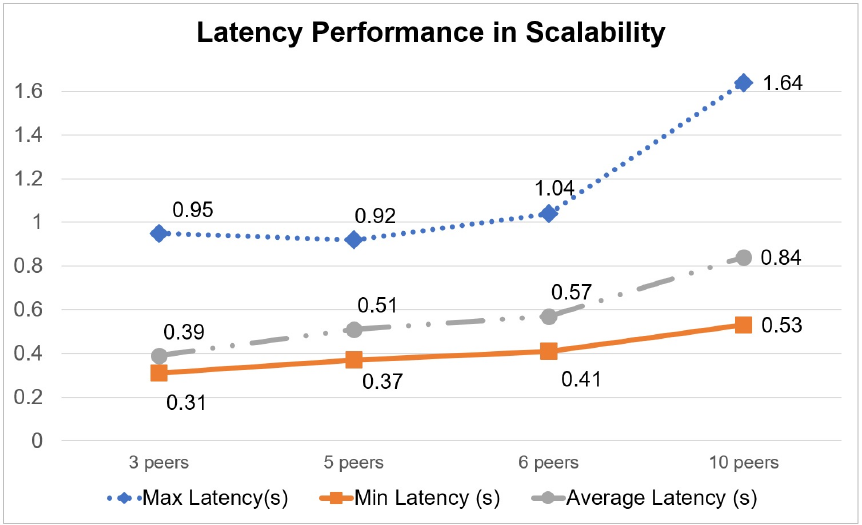}
\caption{Latency of B-DAC in scalability}
\label{fig7}
\end{figure}

To explore the scalability of B-DAC, we additionally deploy and test B-DAC in other four cases including 3 peers, 5 peers, 6 peers, and 10 peers on the same hardware with 8 cores CPU and 32 GB RAM. The scalability performance is measured by sending 1000 requests to the system. The memory consumption and in-out traffic of one peer (peer0) are illustrated in \textbf{Fig.~\ref{fig5}}. Also, the CPU usage of peer0 is depicted in \textbf{Fig.~\ref{fig6}}. The other peers share the same trend with this one. \textbf{Fig.~\ref{fig7}} shows the latency in creating transactions of B-DAC in 4 cases, proving that the scheme of decentralized access control can be efficiently scaled when demands change in the future.

\subsubsection{Security impact of B-DAC to SDN}
Regarding security analysis, various security models like STRIDE, PASTA, Trike, UMLSec, etc., can be applied to identify the security level of a specific system, according to A. Chikhale et al. \cite{c49}. These models define a set of security aspects or characteristics that a system must satisfy to be considered as being secured. Among, STRIDE model \cite{c50} proposed by Microsoft, which is abbreviated for six categories of threats, is considered as the most compatible methodology for classifying security issues for known and unknown attack types. In this work, we also evaluate B-DAC based on the criteria of STRIDE via test scenarios and function analysis.  \textbf{TABLE~\ref{tab7}} summarizes threats and security properties defined in STRIDE and our corresponding experiment scenario to confirm it.

\begin{table}[!b]
\caption{Threats in STRIDE model and testing scenarios}
\begin{center}
\begin{tabular}{|c|c|c|}
\hline
 \textbf{\textit{Property}} & \textbf{\textit{Threat}} & \textbf{\textit{Scenarios}} \\
\hline
Authentication&Spoofing&1, 2\\
Integrity&Tampering&3, 4\\
Non-repudiation&Repudiation&3, 4\\
Confidentiality&Information Disclosure&1, 2, 3, 4\\
Availability&Denial of Service&5\\
Authorization&Elevation of Privilege&3, 4, 6\\
\hline
\end{tabular}
\label{tab7}
\end{center}
\end{table}

However, to have an intuitive understanding, we introduce two typical applications called \emph{MON\_APP} and \emph{FW\_APP} to perform experiments. The former one has not requested any token from the B-DAC system. Meanwhile, the latter has its JWT, and the issued application-controller access token integrated into requests. Their pre-granted permission sets are described in \textbf{TABLE~\ref{tab8}}. These applications send requests to the controller, then receive responses as shown in detail in the following scenarios. Specifically, a sample of the data structure for each verification request is illustrated in \textbf{TABLE~\ref{tab9}}.

\begin{table}[!b]
\caption{Granted permission for OpenFlow applications}
\begin{center}
\begin{tabular}{|c|c|c|}
\hline
 \textbf{\textit{Application}} & \textbf{\textit{Tokens}} & \textbf{\textit{Granted permission}} \\
\hline
\makecell{MON\_APP \\ (appId: app1)}
&None&\makecell{FL\_GET\_SWITCH\_JSON\\FL\_GET\_DEVICE\\FL\_GET\_SINGLE\_SWITCH\\FL\_GET\_LINKS\_JSON\\FL\_GET\_EXERNALLINK\_JSON\\FL\_POST\_ADD\_ACL\\}\\
\hline
\makecell{FW\_APP\\
(appId: app2)
}&\makecell{Access token\\App-controller\\token}&\makecell{FL\_GET\_FW\_RULES\_JSON\\FL\_GET\_FW\_STATUS\_JSON\\FL\_PUT\_ENABLE\_FIREWALL\\FL\_PUT\_DISABLE\_FIREWALL\\FL\_POST\_FIREWALL\_RULE\\FL\_DELETE\_FIREWALL\_RULE\\}\\
\hline
\makecell{MON\_FW\_APP\\
(appId: app3)
}&\makecell{Access token\\App-controller\\token}&\makecell{FL\_GET\_FW\_RULES\_JSON\\FL\_GET\_FW\_STATUS\_JSON\\}\\
\hline
\end{tabular}
\label{tab8}
\end{center}
\end{table}

\begin{table}[!b]
\caption{A data sample of verification request from controller to B-DAC}
\begin{center}
\begin{tabular}{|l|}
\hline
\{ \\
"\$class": "org.blockas.verifyRequest.VerifyingRequest", \\
"url": "/wm/core/switch/", \\
"data": "[]", \\
"tokenId": "pF9LIKdaNhpqSqfAncoQztEObssojZXenU1WqvlC…", \\
"httpMethod": "GET", \\
"permissionId":"FL\_GET\_SINGLE\_SWITCH" \\
\}\\
\hline
\end{tabular}
\end{center}
\label{tab9}
\end{table}

\paragraph{Scenario 1: Authentication to B-DAC}
In this case, we perform the authentication process of an application called \emph{MON\_APP} to the REST API of B-DAC. As mentioned before in \textbf{Section~\ref{sec5.4.1.1}}, an access token and an identity card (wallet) of the application are mandatory for this task. If one of these objects is missing, the authentication is failed. We use \emph{MON\_APP} to query an API called \emph{system\_ping} to get the information about this application. We monitor the response of B-DAC for \emph{MON\_APP} in different cases where the application has not yet authenticated, completed the partial or adequate authentication process. \textbf{TABLE~\ref{tab10}} describes responses of B-DAC when MON\_APP communicates with its REST API in these cases.

\begin{table}[t]
\caption{B-DAC REST API responses in authentication cases }
\begin{center}
\begin{tabular}{|c|c|c|}
\hline
 \textbf{\textit{Cases}} & \textbf{\textit{Description}} & \textbf{\textit{B-DAC response}} \\
\hline
Case 1&\makecell{MON\_APP has no access\\ token}&\makecell{DENY\\
(Authorization required)}\\
\hline
Case 2&\makecell{MON\_APP completes 1 step\\ and receives access token from\\ B-DAC REST API}&\makecell{DENY\\
(Authorization required)}\\
\hline
Case 3&\makecell{MON\_APP completes 2 steps\\
(1)	Received access \\token from B-DAC REST API\\
(2)	Uploaded wallet address\\}&\makecell{ACCEPT\\
(Return app information)
}\\
\hline
\end{tabular}
\label{tab10}
\end{center}
\end{table}

\begin{figure}[!t]
\centering
\includegraphics[width=0.45\textwidth]{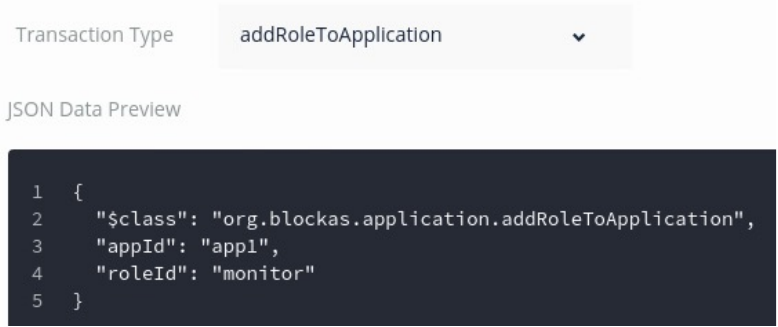}
\caption{Add new permission into an application via blockchain transaction.}
\label{fig9}
\end{figure}

\begin{figure}[!t]
\centering
\includegraphics[width=0.45\textwidth]{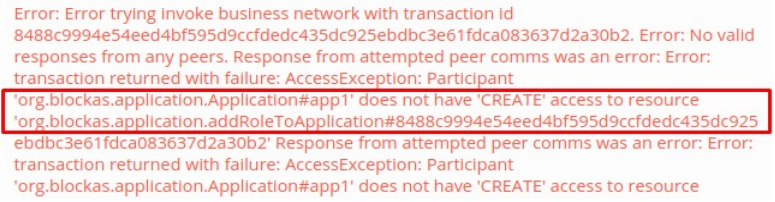}
\caption{B-DAC declines a new permission insertion from application (app1).}
\label{fig10}
\end{figure}

\paragraph{Scenario 2: Authentication between applications and controller}
In B-DAC design, an application needs to utilize the issued app-controller token by administrators when communicating with the controller. In this case, we intend to leverage\emph{ MON\_APP} to create a new ACL rule in Floodlight controller by sending a request to URI \emph{/wm/acl/rules/json}. The required permission for this API is \emph{FL\_POST\_ADD\_ACL}, which has been granted to \emph{MON\_APP}. Therefore, the request should be successfully executed. However, we test the response of B-DAC to \emph{MON\_APP} in the two cases. The first one is an application that does not have a valid app-controller token, while the second case is this app with got issued token from the system. In the point of the controller, we observe results from B-DAC according to these cases, respectively. B-DAC rejects a request if the token is NULL or within the status of NEW and EXPIRED, otherwise, the request is processed as usual after receiving an ACCEPT response.
\paragraph{Scenario 3: Granting permission for application by B-DAC’s Access Control }
A set of permissions is only assigned to an application by administrators in B-DAC. Specifically, the application joining in the B-DAC is restricted with the policy of read-only permission, whereas it is not permitted to upgrade its privileges. \textbf{Fig.~\ref{fig9}} demonstrates a transaction structure that \emph{MON\_APP (app1)} is about to insert a new permission \emph{FL\_GET\_LINKS\_JSON} (belong to the role of monitoring) into its profile. But, B-DAC declines this action, as shown in \textbf{Fig.~\ref{fig10}}. On the contrary, the administrator can add new permissions to this application.

\paragraph{Scenario 4: Application authorization using permissions}
This experiment performs a verification of the authorization when an application requires critical network resources from the controller. The administrator uses a permission called \emph{FL\_PUT\_ENABLE\_FIREWALL} for turning on and turning off the firewall in the Floodlight controller. \emph{FW\_APP} and \emph{MON\_FW\_APP} are used in this scenario, but only \emph{FW\_APP} is designated the relevant permission.

These applications attempt to enable the firewall by sending requests to API \emph{/wm/firewall/module/enable/json}. \textbf{TABLE~\ref{tab11}} shows the response of controller for each application. Moreover, B-DAC also provides the transaction \emph{removePermission} to delete the existing permission from application’s profile.
 
 \begin{table}[!t]
\caption{ Responses from B-DAC for application access}
\begin{center}
\begin{tabular}{|c|c|}
\hline
\textbf{\textit{App}}& \textbf{\textit{B-DAC response}} \\
\hline
FW\_APP & DENY (Unauthorized)\\
\hline
MON\_FW\_APP &  ACCEPT (Firewall status is changed)\\
\hline
\end{tabular}
\label{tab11}
\end{center}
\end{table}

\begin{table*}[t]
\caption{Scenario and rule samples for evaluation of conflict detection}
 \centering
 \begin{tabular}{|c|c|c|c|c|c|c|c|}
\hline
\textbf{\textit{Subscenario}}& \textbf{\textit{Rule}} & \textbf{\textit{Protocol}} & \textbf{\textit{Source}} & \textbf{\textit{Destination}} & \textbf{\textit{Priority}} & \textbf{\textit{Action}} & \textbf{\textit{B-DAC Response}}\\
\hline
S1 & \makecell{r1 \\ r2} & \makecell{TCP \\ TCP} & \makecell{10.0.0.0/24 \\ 10.0.0.0/32} & \makecell{10.0.0.0/24 \\ 10.0.0.2/32} & \makecell{51 \\ 50} & \makecell{ALLOW \\ ALLOW} & \makecell{SUCCESS \\ CONFLICT (Redundancy)} \\
\hline

S2 & \makecell{r3 \\ r4} & \makecell{ICMP \\ ICMP} & \makecell{10.0.0.0/24 \\ 10.0.0.0/24} & \makecell{10.0.0.0/24 \\ 10.0.0.2/32} & \makecell{52 \\ 51} & \makecell{ALLOW \\ DENY} & \makecell{SUCCESS \\ CONFLICT (Shadowing)} \\
\hline

S3 & \makecell{r5 \\ r6} & \makecell{TCP \\ TCP} & \makecell{10.0.0.0/24 \\ 10.0.0.0/32} & \makecell{10.0.0.0/24 \\ 10.0.0.2/32} & \makecell{50 \\ 50} & \makecell{ALLOW \\ DENY} & \makecell{SUCCESS \\ CONFLICT (Correlation)} \\
\hline

S4 & \makecell{r7 \\ r8} & \makecell{UDP \\ UDP} & \makecell{10.0.0.1/32 \\ 10.0.0.0/24} & \makecell{10.0.0.2/32 \\ 10.0.0.0/24} & \makecell{52 \\ 53} & \makecell{ALLOW \\ DENY} & \makecell{SUCCESS \\ CONFLICT (Generalization)} \\
\hline

S5 & \makecell{r9 \\ r10} & \makecell{TCP \\ TCP} & \makecell{10.0.0.0/28 \\ 10.0.0.1/32} & \makecell{10.0.0.0/28 \\ 10.0.0.0/24} & \makecell{51 \\ 55} & \makecell{DROP \\ DROP} & \makecell{SUCCESS \\ CONFLICT (Overlap)} \\
\hline

S6 & \makecell{r11 \\ r12 \\ 13} & \makecell{TCP \\ TCP \\ ICMP} & \makecell{10.0.0.0/28 \\ 10.0.0.16/29 \\ 10.0.0.1/32} & \makecell{10.0.0.0/28 \\ 10.0.0.24/29 \\ 10.0.0.2/32} & \makecell{51 \\ 52 \\ 55} & \makecell{DENY \\ ALLOW \\ DENY} & \makecell{SUCCESS \\ SUCCESS \\ SUCCESS} \\
\hline

\end{tabular}
\label{tab13}
\end{table*}
\paragraph{Scenario 5: Resource starvation prevention}
To guarantee the service quality and avoid resource exhaustion at the controller, we define a limit of accessing rate to Northbound API for all applications. Specifically, we set the limit rate to 1200 requests each 30 seconds for each application. This rate is determined by management policy from the administrator. In our design, each application cannot overdo the quota assigned in its profile. As a result, a request which belongs to an over-using application is rejected immediately to further process by B-DAC and SDN network. The mechanism of quota-based resource allocation helps the controller preserve the availability against torrential requests aiming to disrupt the controller operation.
\paragraph{Scenario 6: Flow rules conflict prevention from authorized application}

To evaluate the \emph{Flow rule conflict detector} module of B-DAC, this scenario uses an application called \emph{MON\_ACL\_APP}, which has successfully passed the authentication and authorization phases. This application interacts with the supported ACL API on Floodlight controller and allows users to define access control rules that they want to apply to the network. These rules are then used to create corresponding flow rules to install on SDN devices. We create and test rules in both conflicted and normal cases to evaluate the effectiveness of the conflict detecting algorithm. Confliction in test rules is ranged in the aforementioned five types.

\begin{figure}[t]
\centering
\includegraphics[width=0.45\textwidth]{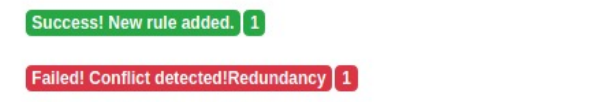}
\caption{Notifications from controller for ACL rule adding requests.}
\label{fig11}
\end{figure}

Generally, there are six sub-scenarios in this evaluation, including five different ones for each type of confliction and another case of adding conflict-free rules. In each sub-scenario, at least two ACL-form defined rules are sent to the controller via the URL of ACL API, which is \emph{/wm/acl/rules/json}. In \emph{MON\_ACL\_APP} application, the response of B-DAC for requests of adding rule is also recorded. As shown in \textbf{Fig.~\ref{fig11}}, where the green label (the first line) notifies a successful one and the red lines (the second line) are used to mention errors.
 
\textbf{TABLE~\ref{tab13}} lists rules used in each sub-scenario, and summaries their main parameters which are the concern of conflict detecting algorithm. The response, as well as the conflict detection results of B-DAC, are also given correspondingly in the last column.

\subsubsection{Evaluating B-DAC-enabled controller overhead}
To evaluate the extra overhead of B-DAC enabled controller, we consider the metrics of CPU and memory consumed when performing a various number of application requests. Notably, the experiment results in this section give a comparison of original Floodlight and B-DAC-enabled Floodlight controller. The overhead of SDN controller which integrates with B-DAC consists of two tasks: the request processing on Floodlight controller and blockchain transaction execution. Because the processing delay overhead is affected by different response results (ACCEPT or DENY), we set all API requests from applications to be benign for ensuring the fairness of experiments. This means that unauthorized requests will be promptly rejected by B-DAC with low latency than that of benign requests.

To start with, we perform total 1000 requests (reading statistics of an OpenFlow switch from controller's API) from one application, called \textit{app0} to B-DAC-integrated controller. The average response time is around 0.01735 seconds. Meanwhile, CPU consumption to process 1000 requests increases by approximately 3.2\% comparing to the background running of the B-DAC-enabled Floodlight controller. In comparison to the original Floodlight (2.4\%), this is a slight growth because B-DAC-enabled Floodlight must analyze permissions and wait for the response from B-DAC.
In another case, we perform experiments with 6 applications (\textit{app0, app1, app2, app3, app4, app5}). Therein, each of them sends 500 requests to produce 3000 requests in total to SDN controller. In a third case, 6 applications above are used to send requests to a controller with 1000 requests per application. As a result, with a total of 6000 different requests processed, the average response time slightly climbs to 0.02123 seconds and CPU load increases by 6.3\%. Thus, we can see that if the number of requests raises 6 times, the B-DAC-integrated controller witnesses an increase of 22.3\% in response time. Besides, our B-DAC-enabled controller consumes roughly 5.2 MB, 14.7 MB, 32.5 MB of memory for 1000, 3000, 6000 requests, respectively. The measurement results on overhead metrics are illustrated in \textbf{TABLE~\ref{tab14}}. These results indicate that the computational overhead of a B-DAC-integrated controller is not significant. 

\begin{table*}[t]
\caption{B-DAC-integrated Floodlight controller's overhead performance}
\centering
\begin{tabular}{|c|c|c|c|c|c|c|}
\hline
\#                                                            & \multicolumn{2}{c|}{\begin{tabular}[c]{@{}c@{}}Average \\ delay time (s)\end{tabular}}                                                                                            & \multicolumn{2}{c|}{\begin{tabular}[c]{@{}c@{}}RAM usage \\ for requests (MB)\end{tabular}}                                            & \multicolumn{2}{c|}{\begin{tabular}[c]{@{}c@{}}CPU usage \\ for requests (\%)\end{tabular}}                                            \\ \hline
\begin{tabular}[c]{@{}c@{}}Number \\ of requests\end{tabular} & \multicolumn{1}{l|}{\begin{tabular}[c]{@{}l@{}}B-DAC\\ -enabled \\ Floodlight\end{tabular}} & \multicolumn{1}{l|}{\begin{tabular}[c]{@{}l@{}}Original \\ Floodlight\end{tabular}} & \begin{tabular}[c]{@{}c@{}}B-DAC\\ -enabled\\ Floodlight\end{tabular} & \begin{tabular}[c]{@{}c@{}}Original \\ Floodlight\end{tabular} & \begin{tabular}[c]{@{}c@{}}B-DAC\\ -enabled\\ Floodlight\end{tabular} & \begin{tabular}[c]{@{}c@{}}Original \\ Floodlight\end{tabular} \\ \hline
1000                                                          & 0.01735                                                                                     & 0.0067                                                                              & 5.2                                                                   & 4.6                                                            & 3.2\%                                                                 & 2.4\%                                                          \\ \hline
3000                                                          & 0.01974                                                                                     & 0.0075                                                                              & 14.7                                                                  & 12.8                                                           & 4.3\%                                                                 & 3.4\%                                                          \\ \hline
6000                                                          & 0.02123                                                                                     & 0.0089                                                                              & 32.5                                                                  & 29.2                                                           & 6.3\%                                                                 & 4.7\%                                                          \\ \hline
\end{tabular}
\label{tab14}

\end{table*}

To summarize, to show the effectiveness in preventing attacks starting from malicious applications, we launch some attack scenarios to check whether it can bypass the access control to consume network resources or not, as depicted in the above experiments. The result demonstrates that B-DAC can support a security-enhanced SDN controller at the Northbound interface due to corresponding policies defined by network administrators. As a result, one application is only permitted to consume resources belonging to its profile. All undeclared action is strictly denied. This mechanism also prohibits privileges escalation requests stemming from a compromised application, then promptly rejects or blocks its functions in the network.
\section{Related works}
\label{sec7}
SDN has been on its way to become an emerging alternative solution to modern networks. Therefore, the problem of enhancing security in SDN is gaining more and more attention from research experts. In the context of the Northbound interface, several solutions for secured communication between applications and SDN controllers have been proposed. Those works both aim to provide authentication and control mechanisms to prevent controllers from being compromised for malicious activities. Our work is inspired by prior works in SDN security and vulnerability analysis techniques, particularly regarding NBI security, as follows.

The first attempts to protect controllers seem to be to enable an authentication and authorization mechanism to restrict unauthorized access to this component. 

To achieve this goal, Philip Porras et al. developed a kernel solution called FortNOX \cite{c31}. Their work is designed to deal with the intrusion of malicious OF apps to the controller and unauthorized flow rule installation. It is then also extended to another version named SE-Floodlight \cite{c32}. Both solutions classify applications into three groups and use a role-based authentication mechanism. However, in their work, the controller still takes the responsibility of access control and conflict verification. It could be a bad effect on the performance of the controller when it must process continuously an enormous volume of requests. For instance, the response time of requests can increase. Even worse, the controller may get overloaded and become unavailable for normal network controls. Similarly, SAIDE proposed by Tao Hu et al. \cite{c29} also has the controller taken responsibility for detecting and eliminating application interferences in SDN. Clearly, this solution is completely controller-dependent.

In another work, A.L. Aliyu et al. \cite{c33} proposed a trust management framework for the access of OpenFlow applications to the controller. Each OpenFlow application is granted with specific permissions, which are Read, Write, Notify, and SystemCall. Using this property, this framework aims to ensure no OF app can overstep its granted permission to perform unprivileged actions. Moreover, the trust of the controller on applications is also established and monitored by this framework. Basing on the application's behaviors in SDN, its corresponding trust value is periodically reviewed and updated.

Also considering trust as an important factor, B. Isong et al. \cite{c34} used the trust level of an OF app as its “access card” to SDN controller. An application must satisfy a required trust level to be able to communicate with the controller and request network resources. This approach creates a trust matrix of applications corresponding to their identities used for effective resource management in SDN.

Besides, a solution called ControllerSEPA \cite{c24} utilized a repacking service to manage the transferred data from the controller to OF app. In other words, it works as an intermediate communication layer between application and control planes for intercepting and analyzing traffic. It also provides services of AAA based on the features and granted permissions of a specific OF app. Moreover, these applications interact with the controller via a TLS-enabled Northbound interface to ensure secured communication. An advantage of this work is being developed as a plug-in model placed outside the controller. This design makes no significant effect on the overall performance of the SDN control plane.

Likewise, Tao Hu et al. presented SEAPP \cite{c25}, a secure application management framework for SDN-aware cloud networks. To prevent malicious applications from abusing REST APIs to launch hostile attacks, they designed an access control scheme on the Northbound interface in SDN. On a more detailed level, network administrators have OpenFlow apps registered with permission manifests. In later communications between applications and the SDN controller, SEAPP compares the permissions in the application's requests with declared permissions to allow access to network resources. However, this access control framework is susceptible to denial of services attacks. Even worse, man-in-the-middle (MITM), spoofing can be new vulnerabilities of SEAPP if the policy database is exploited.

In other work, B. Toshniwal et al. proposed BEAM \cite{c35} - a concept of a behavior-based access control mechanism for SDN applications. This solution aims to grant and update applications’ permission dynamically and periodically by analyzing their activities on SDN. The activities are verified using their corresponding logs, then are used to define behaviors with many metrics like \emph{packet\_in\_rate}, \emph{flow\_injection\_rate}… Though this mechanism is just in theory, its practical implementation can be promising to protect the SDN controller from malicious applications.

Despite the aforementioned studies have expressed their strength in protecting controllers, they still encounter many aside problems such as performance reduction \cite{c31} \cite{c32} or single point of failure due to the bridged position \cite{c24}. Moreover, most traditional authentication and authorization systems use username-password pairs and other information saved in a regular database as the identities. However, there is a fact that these databases can be illegally modified. In this case, the outstanding immutability characteristic of blockchain makes it become a promising alternative solution. More specifically, it can ensure data integrity and enhance the system's availability by the decentralized nature.

Regarding the SDN context, blockchain has had its very first applications in enhancing the security of this architecture. A survey by Wenjuan Li et al. \cite{c36} introduced a generic framework of Blockchain-based SDN. They also have an overview of challenges and solutions in combining these two technologies. Though many security issues may still exist, blockchain and SDN can complete each other in many practical scenarios to provide more secured architectures.

In a vehicular environment, called the Internet of Vehicles (IoV), security and privacy-preserving should be a major concern due to involving transport systems and autonomous cars. With the rapid development of vehicle networking in scalability, SDN is also applied to improve the IoV network management. In such SDN-based IoV (SD-IoV) systems, the lack of authentication, authorization and trust management of the applications also brings many security challenges like SDN architecture. To this end, Mendiboure L. et al. \cite{c26} proposed a blockchain-based approach to provide trust establishment between SDN controllers and network applications. In this system, a smart contract is used to store the information necessary for the application authentication process in an identity card. All transactions and exchanged data are kept in the blockchain network when controllers and network applications are mutually authenticated. This principle makes the SDN-IoV controller more secure from malicious IoV applications. Nonetheless, their work is only a conceptual model, not to be implemented and evaluated in experimental scenarios.

Moreover, Zhedan Shao et al. \cite{c37} built a distributed database using Blockchain to monitor operations of the controllers. This database maintains a list of system actions in each controller, which provides a strong and unmodified data store. Therein, the authors use the SPBFT as the consensus algorithm to enable parallel message transfer to speed up the exchanged data processing.

Not limited to application authentication on Northbound interface, Blockchain is also used to guarantee SDN security from the view of data plane. In a research \cite{c38}, the authors introduced a Blockchain-based authentication mechanism in data plane of SDN. Each SDN subnet has a private chain formed inside, and its data are stored in blockchain for integrity. A new node willing to access SDN must pass the authentication of the controller and a randomly chosen authentication node. Then, it can receive keys from the security gateway of the subnet. These keys are used to sign and decrypt the transaction, which is a process to identify each node in the network.

Along with the authentication-related issues, SDN also requires attention in protecting the controller from misuse operations of legal or seem-to-be-safe applications. As flow rules play a key role in forwarding packets, the attacker can take advantage of these rules to command the network. As a proof, the authors of SRV \cite{c39} proposed an attack model using malicious flow rules. Then, the impact of the attack is analyzed to evaluate the risk level. Hence, problems in flow rules, such as rule validation and verification, have become other interesting topics in SDN security.

A work \cite{c28} presented a scheme called PERM-GUARD as a flow rules validation module in Floodlight controller in SDN. This scheme introduces a new model to verify and authenticate the validity of the flow rule in SDN. In this work, every application intending to communicate with the controller must register themselves to be recognized via unique ID and identity-based signature. To decide whether flow rules should be installed on switches, PERM-GUARD ensures that two following criteria are both valid. The first one is that the granted permissions of the application have met the required ones of the rule, and the other is the validity of the producer of that rule.

Besides, S. Wang et al. \cite{c38} improved the integrity of flow rule by storing it on the blockchain network whenever switches receive and update their flow tables. So, that, SDN network nodes can periodically query the correct flow table information and change their local ones if it is inconsistent. Moreover, S. Pisharody \cite{c40} categorized conflicts into various types and proposed an algorithm to detect and solve the conflicts. The effectiveness of this work has been proved via multiple testing scenarios of rule confliction in SDN.

In the urgent need for securing the Northbound interface, our work is motivated by not only existing problems in the above approaches, but also their strength and potentially extend capabilities. In this paper, we aim to propose a Blockchain-based authentication and control framework for the SDN Northbound interface, called B-DAC. To the best of our knowledge, B-DAC is the first approach that leverages blockchain with a prototype implementation to provide a decentralized and fine-grained filter to control network applications in SDN. Our work integrates the strength of current solutions. At first, our framework has a proactive approach like a plugin model in ControllerSEPA’s idea. This design enables authentication and access control tasks to be performed independently with the controller so that it ensures the SDN network brain’s performance. Moreover, all the data used in the operations of the framework is stored in blockchain, where it is impossible to edit or remove illegally. In addition, the framework utilizes certificates as the identities of its managed SDN elements, such as controller and OpenFlow (OF) application. In the case of OpenFlow applications, its behaviors are also paid attention to consider the trustworthiness of an application from B-DAC. This trust degree is measured and updated according to the application behaviors in SDN. Along with common authentication, authorization, and accounting capabilities, B-DAC can also validate incoming requests of rule installation to ensure the proper operation of the SDN controller as well as the entire network.
\section{Conclusion and Future Work}
\label{sec8}
In this paper, we propose B-DAC – a blockchain-based access control for the Northbound interface in SDN. It is the decentralized access control framework with prototype implementation using Hyperledger Fabric blockchain approach for securing SDN controller. Our framework aims to secure the interaction between SDN controllers and network applications by utilizing the underlying characteristics of the blockchain. Specifically, B-DAC is designed to achieve controller-independent, application-transparency, and strict and decentralized access control. It ensures that all communications from any application to the controller are always verified before network resources are consumed. With this approach, it is infeasible for hackers to create a fake entity to launch malicious attacks targeting the controller-application channel. Additionally, we also enforce the fine-grained observation of OpenFlow applications in B-DAC by security policy generation. It helps network administrators prevent the problems of unauthorized access, privilege escalation, and network resources exhaustion during resource allocation and management. Based on its implementation and evaluation in many experimental scenarios, we conclude that B-DAC obtains accurate and fine-grained policy enforcement for Northbound interface without significantly degrading network performance. Also, we claim that our initial architecture concentration is security, not performance, due to its trade-off.
Nevertheless, our work has some limitations that can be improved in the future. Flow rules conflicts in the SDN network are still detected and prevented directly on a specific controller (i.e., Floodlight), although other modules of B-DAC are independent of the SDN controller. In the future, we can relocate the validator of the flow rule outside the controller. 

Also, to relieve the depression of administrators when creating permission in the policy definition phase, we can adapt natural language processing (NLP) techniques to automatically generate permissions for B-DAC. In this case, an example solution like VOGUE \cite{c21} can be considered as a potential approach to integrate with B-DAC for the mentioned purpose. This may dramatically improve the workload of management in the network. Moreover, when administrators verify the token, it is vital to make the phase of issuing tokens automatically to replace human interventions. 
When it comes to the distributed controller model, B-DAC only uses one controller to perform all decision-making for the forwarding plane in SDN. Thus, the implementation of multiple controllers for B-DAC will be considered to evaluate the effectiveness of our design in complicated contexts. In addition, B-DAC can also apply in the context of IoT devices authentication to identify and trust each other for data exchange.

 Furthermore, system performance is also one of our considerations. Transaction speed and throughput in the blockchain network are important factors in this improvement journey. This can be tackled by utilizing such an approach like FastFabric \cite{c51}, a plug-and-play approach on Hyperledger Fabric, aiming to reduce computation and I/O overhead during transaction ordering and validation to significantly improve throughput. Also, we can deploy blockchain nodes on multiple hosts with robust hardware requirements to increase the performance of the blockchain network. In addition, instead of using Hyperledger Composer, we will implement the chaincode with Hyperledger Fabric SDK. This method allows us more deeply intervene in the provided blockchain APIs. At that time, it will optimize the number of transactions of the blockchain network, which can be up to thousands of transactions per second. The better results of these metrics can provide the better performance of keeping data immutably for blockchain-based solutions in SDN when witnessing the data explosion of network traffic.
\section*{Acknowledgment}
Phan The Duy was funded by Vingroup Joint Stock Company and supported by the Domestic Master/ PhD Scholarship Programme of Vingroup Innovation Foundation (VINIF), Vingroup Big Data Institute (VINBIGDATA), code VINIF.2020.TS.138.

The authors would like to express the special thanks of gratitude to all astoundingly supportive members in E8.1 room as well as friends, teachers who gave us the opportunity to do this wonderful project.
\ifCLASSOPTIONcaptionsoff
  \newpage
\fi



%

%

\begin{IEEEbiography}[{\includegraphics[width=1in,height=1.25in,clip,keepaspectratio]{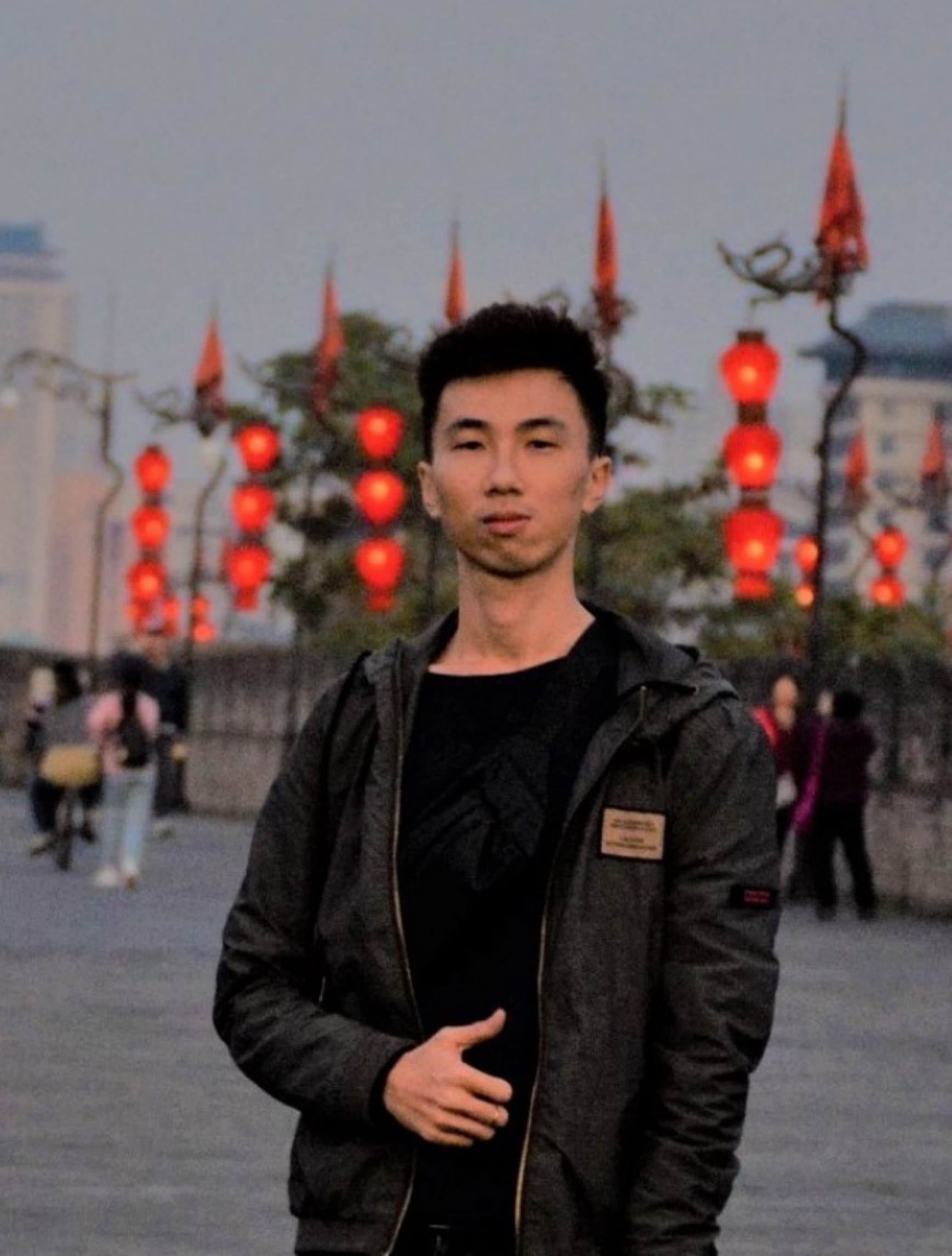}}]{Phan The Duy}
received the B. Eng. and M.Sc. degrees in Software Engineering and Information Technology from the University of Information Technology (UIT), Vietnam National University Ho Chi Minh City (VNU-HCM), Hochiminh City, Vietnam in 2013 and 2016, respectively. Currently, he is pursuing a Ph.D. degree major in Information Technology, specialized in Cybersecurity at UIT, Hochiminh City, Vietnam. He also works as a researcher member in Information Security Laboratory (InSecLab), UIT-VNU-HCM after 5 years in the industry, where he devised several security-enhanced and large-scale teleconference systems. His research interests include information security \& privacy, Software-Defined Networking security, digital forensics, adversarial attacks, generative adversarial networks (GANs), machine learning-based security solution and blockchain.
\end{IEEEbiography}
\begin{IEEEbiography}[{\includegraphics[width=1in,height=1.25in,clip,keepaspectratio]{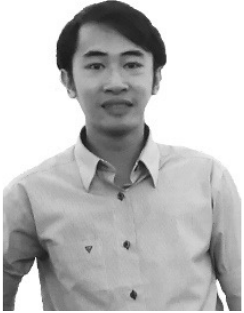}}]{Hien Do Hoang}
received B.E degree in Networking and Communication from the University of Information Technology (UIT), Vietnam National University Ho Chi Minh City (VNU-HCM), Vietnam in 2017. From 2017 to 2018, he worked for a security and network company. He also received the M.Sc. degree in Information Technology in 2020. Currently, he works as a researcher member in Information Security Lab (InSecLab) in UIT, VNU-HCM. His research interests are Software-defined Networking, system and network security, cloud computing and blockchain.
\end{IEEEbiography}
\begin{IEEEbiography}[{\includegraphics[width=1in,height=1.25in,clip,keepaspectratio]{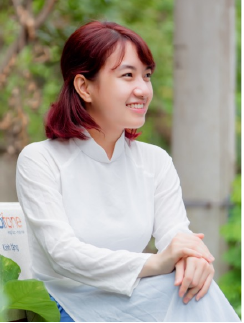}}]{Do Thi Thu Hien}
received the B. Eng. degree in Information Security from the University of Information Technology (UIT), Vietnam National University Ho Chi Minh city in 2017. She received the M.Sc. degree in Information Technology in 2020. From 2017 till now, she works as a member of a research group at the Information Security Laboratory (InSecLab) in UIT. Her research interests are Information security \& privacy, Software-defined Networking, and its related security-focused problems.
\end{IEEEbiography}
\begin{IEEEbiography}[{\includegraphics[width=1in,height=1.25in,clip,keepaspectratio]{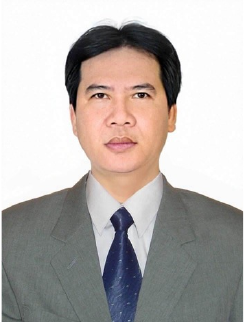}}]{Anh Gia-Tuan Nguyen}
obtained his bachelor’s degree in Information Technology from the University of Natural Sciences, Vietnam National University Ho Chi Minh City (VNU-HCM), Hochiminh City in 1995. Then, he pursued his M.Sc. degree in Information Technology from this university in 1998.  He completed his Ph.D. thesis in Information Technology from the University of Natural Sciences of Hochiminh City in 2013. He is now a lecturer at the University of Information Technology, Vietnam National University Ho Chi Minh City (UIT-VNU-HCM). His main research interests include information security, 3D/4D GIS, mapping application and modern databases.
\end{IEEEbiography}
\begin{IEEEbiography}[{\includegraphics[width=1in,height=1.25in,clip,keepaspectratio]{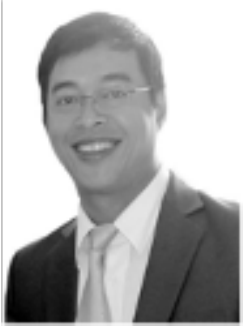}}]{Van-Hau Pham}
obtained his bachelor’s degree in computer science from the University of Natural Sciences of Hochiminh City in 1998. He pursued his master’s degree in Computer Science from the Institut de la Francophonie pour l’Informatique (IFI) in Vietnam from 2002 to 2004. Then he did his internship and worked as a full-time research engineer in France for 2 years. He then persuaded his Ph.D. thesis on network security under the direction of Professor Marc Dacier from 2005 to 2009. He is now a lecturer at the University of Information Technology, Vietnam National University Ho Chi Minh City (UIT-VNU-HCM), Hochiminh City, Vietnam. His main research interests include network security, system security, mobile security, and cloud computing.
\end{IEEEbiography}
\balance




\end{document}